%% file: qbf-gallery-report.tex
\pdfoutput=1
\documentclass[preprint]{elsarticle}
\usepackage{amsmath, amsthm, amssymb}
\usepackage[utf8]{inputenc}
\usepackage{graphicx} 
\usepackage{latexsym}
\usepackage{geometry}
\usepackage{url}
\usepackage{color}
\usepackage{rotating}
\usepackage{multirow}
\usepackage{xspace}
\usepackage{multicol}
\usepackage{alltt}
\usepackage{ifthen}
\usepackage{tikz}
\usepackage{epstopdf}

\theoremstyle{definition}

\theoremstyle{remark}

\newboolean{showappendix}
\setboolean{showappendix}{true}

\input{macros.tex}

\begin{document}
\begin{frontmatter}

\title{The QBF Gallery: Behind the Scenes\tnoteref{thanksfootnote}}
\tnotetext[thanksfootnote]{Supported by the Austrian
    Science Fund (FWF) under grants S11408-N23 and  S11409-N23. \textbf{This
      article  will appear in \emph{Artificial Intelligence}, Elsevier, 2016.}}

\author[lonsing]{Florian Lonsing}
\author[seidl]{Martina Seidl}
\author[vangelder]{Allen Van Gelder}

\address[lonsing]{Vienna University of Technology, Austria}
\address[seidl]{Johannes Kepler University, Linz, Austria}
\address[vangelder]{University of California at Santa Cruz, USA}

\begin{abstract}
Over the last few years, much 
progress has been made in the theory and practice of solving 
quantified Boolean formulas (QBF). Novel solvers have 
been presented that either successfully enhance established techniques 
or implement novel solving paradigms. Powerful preprocessors have 
been realized that tune the encoding of a formula to make it 
easier to solve. Frameworks 
for certification and solution extraction emerged that allow for a detailed
interpretation of a QBF solver's results, and new types 
of QBF encodings were presented for various application problems. 

To capture these developments
the \emph{QBF Gallery} was established in 2013. The QBF Gallery aims 
at  providing a forum
to  assess QBF tools and to collect new, expressive benchmarks
that allow for documenting the status quo and that indicate 
promising research directions. These 
benchmarks became the basis for the experiments 
conducted in the context of the QBF Gallery 2013 and 
follow-up evaluations. 
In this paper, we report on the setup of the QBF Gallery. To this end, 
we conducted numerous experiments which allowed us not only to assess 
the quality of the tools, but also the quality of the benchmarks. 
\end{abstract}


\end{frontmatter}

%%%%%%%%%%%%%%%%%%%%%%%%%%%%%%%%%%%%%%%%%%%%%%%%%%%%%%%%%%%%%%%%%%%%%%%%%%%%%%%%
%%%%%%%%%%%%%%%%%%%%%%%%%%%%%%%%%%%%%%%%%%%%%%%%%%%%%%%%%%%%%%%%%%%%%%%%%%%%%%%%

\section{Introduction}
\label{intro}

Quantified Boolean formulas (QBF)~\cite{DBLP:series/faia/BuningB09} 
provide a powerful framework 
for encoding any application problem located in the complexity class
of PSPACE. Many important verification problems like 
bounded model checking~\cite{DBLP:journals/entcs/JussilaB07} or artificial intelligence tasks like
conformance planning~\cite{confplanning14} can be efficiently encoded 
as QBF (cf.~\cite{jsat:benedetti08} for a survey). The use of
existential and universal quantifiers in QBF potentially allows for
encodings which are exponentially more succinct than encodings in
propositional logic (SAT). 
 Given the success story of 
SAT solving~\cite{DBLP:series/faia/2009-185}, much emphasis
and efforts have been spent in repeating this success story for QBF,
with the aim to avoid the space explosion inherent in 
SAT encodings. So far, QBF based-technologies have not yet  
reached the mature state of modern SAT-based technology, 
but nevertheless continuous progress can be observed. 

Recently, several novel approaches  emerged
 ranging from innovative solving techniques 
to effective preprocessing and new encodings of application problems. 
A major breakthrough has been achieved by solving the 
long open problem of calculating certificates for a solver's  
result, leading to elegant approaches based on the analysis of 
resolution proofs~\cite{DBLP:journals/fmsd/BalabanovJ12,DBLP:conf/ijcai/GoultiaevaGB11}, to name one example.  
Such advancements are often distributed over
multiple publications and implemented in different tools with evaluations
performed within different infrastructures,
which makes them hard to compare. Due to this heterogeneity, QBF 
certainly has some entrance barrier for potential contributors and users. 
Therefore, we decided to set up the QBF Gallery
as annual or biannual event,
which is open to all 
advancements in the field of QBF research. 

The first edition of the QBF Gallery was organized in 2013. 
We invited the 
QBF research community
to contribute ideas on what kind of evaluations would be interesting,
given a common infrastructure to perform experiments. 
The QBF Gallery 2013~\cite{qbfgallery2013} 
was a non-competitive, community-driven event affiliated 
with the \emph{First International Workshop on Quantified Boolean 
Formulas (QBF 2013)}\footnote{\url{http://fmv.jku.at/qbf2013/}}. 
The overall goal was to 
evaluate the state-of-the-art of QBF-related technologies. 
This strongly distinguishes the QBF Gallery 2013 event
from previous competitions~\cite{DBLP:conf/sat/PeschieraPTBKL10,qbfevalr2}, 
where the main focus was set on the 
competitive comparison of solvers with the goal to crown winners.
In the QBF Gallery 2013, we abstained from this competitive spirit.
We were interested in performing comprehensive experiments 
that allow us to better understand the benefits and drawbacks of 
different techniques. We did 
not organize a competition in a traditional sense,
so we 
awarded no prizes. Instead, we collected and analyzed 
data obtained during numerous experimental runs.
The participants were immediately provided with all the results and 
their feedback was considered for follow-up experiments. 
Furthermore, in the case of discrepancies in the solving results, 
we immediately informed the respective participants who could then 
submit a fix and continue to participate without any consequences. 
Events like the QBF Gallery are important to give an overview on the 
state of the art and to provide a common forum for watching the 
progress in a research community. For potential users, 
the QBF Gallery should provide an easy entrance into QBF technology by collecting 
current research results manifesting in tools. 

We set up four different showcases for the QBF Gallery 2013. 
The four showcases are (1) solving, (2) preprocessing, 
(3) applications, and (4) certification. 
The \emph{solving showcase} evaluates and compares different solvers
in various scenarios in order to understand the solvers' suitability 
for given benchmarks. Naturally, solving also plays an important 
role in the other showcases.
In the \emph{preprocessing showcase} we were interested in studying the 
impact of individual and combined preprocessors 
on the behavior of the solvers. Recently published 
encodings of application problems,
and the ability of the solvers to handle those were studied in 
the dedicated \emph{application showcase}. Here, we considered only newly 
committed benchmarks.
Finally, the \emph{certification showcase} was dedicated to the evaluation 
of trace producing solvers, and the evaluation of the performance of the 
certification frameworks.
Obviously, the 
showcases are strongly related and results from one showcase might
also be of relevance for the other showcases. 
However, the different showcases allowed us to focus on different 
aspects of the solving process. 

One piece of feedback we received several times for the organization of the 
QBF Gallery 2013 was that some important aspect is missing: 
the competition. Besides the scientific insights, research challenges and 
documentation of the state of the art, one motivation in participating 
in a competition is the fun factor and the direct comparison with competitors.
Therefore, the 
participants asked for a competitive setup where prizes are awarded 
to the best solvers. As a follow-up event of the QBF Gallery 2013, we therefore organized the QBF Gallery 2014
as a traditional solver competition in the context of the FLoC 
Olympic Games\footnote{\url{http://vsl2014.at/olympics/}}, awarding different medals to the best performing 
solvers. For the benchmarks, we reused variants of the sets 
established during the QBF Gallery 2013, which were available 
to all participants.

In this paper, 
we take a look behind the scenes of the QBF Gallery 2013 
and we report on the experiments which yield the basis for the 
conducted tool evaluations. 
To this end, we first summarize the participating systems and
the evaluated benchmarks submitted to the  QBF Gallery 2013 
in Section~\ref{sec:part}. We describe the four different 
showcases that we considered in our experiments: the  solving showcase
is presented in Section~\ref{sec:solving}, followed by the preprocessing 
showcase in Section~\ref{sec:prepro}. In Section~\ref{sec:apps}, 
we report on the application showcase, and finally, in Section~\ref{sec:cert}
we give a short summary on the certification showcase. 
In Section~\ref{sec:conclusion}, we conclude this paper with a short summary 
of the QBF Gallery 2014, which was organized as a competition in the 
context of the FLoC Olympic games. Then we shortly discuss insights gained 
from the organization of the QBF Gallery and conclude with lessons learned.

\section{Setup of the QBF Gallery 2013}
In early 2013, we invited the QBF research community to participate in 
the first edition of the QBF Gallery by contributing ideas, tools, and benchmarks. 
Overall, 23 contributors from eight countries 
provided their tools for experiments. 
The submissions included $15$ solvers for QBFs in conjunctive normal form (CNF), one non-CNF solver, 
three 2-QBF solvers,\footnote
{``2-QBF'' means two quantification levels, it does not mean binary clauses.}
four preprocessors, two certification tools, and five new benchmark sets.
Besides the newly submitted formulas we additionally considered more 
than 7,000 formulas provided by the QBFLIB~\cite{qbflib},
 the community portal of 
QBF researchers. 
With these artifacts, we performed more than 114,000 runs in over 11,000 
CPU hours. Details on the used infrastructure are given with the 
description of the different experiments.
Benchmarks and log files of the runs are available at the website of the QBF Gallery 2013~\cite{qbfgallery2013}.

At the QBF Gallery event of 2013,  we
focused on general aspects of tools in the context of QBFs and not 
only on runtime performance. We set up four 
showcases where we addressed  typical usage scenarios such as 
solving, preprocessing,
novel applications, and strategies/certificates. 
We tried to identify trends and to gain
insights into the performance of the tools. This is very different from
previous QBFEVALs, which focused mainly on the competitive aspects 
in terms of solving performance. On purpose, we decided to use the 
simplest possible performance metrics like number of solved formulas, 
average and total runtimes. These simple metrics were sufficient for our goal of 
understanding how the different systems perform on different benchmark
sets. However, in a competitive setting other metrics (cf.\ for example~\cite{DBLP:conf/sat/Gelder11})might have been more adequate.   

%%%%%%%%%%%%%%%%%%%%%%%%%%%%%%%%%%%%%%%%%%%%%%%%%%%%%%%%%%%%%%%%%%%%%%%%%%%%%%%%
%%%%%%%%%%%%%%%%%%%%%%%%%%%%%%%%%%%%%%%%%%%%%%%%%%%%%%%%%%%%%%%%%%%%%%%%%%%%%%%%

\subsection{Participating Systems and Benchmarks} \label{sec_participant_info}
\label{sec:part}

In this section, we give an overview of the submitted tools and the 
benchmarks used in the experiments. We provide  
references to related literature describing the internals of each 
solvers.
The benchmarks are available at the QBF Gallery 2013 
website~\cite{qbfgallery2013}.

\begin{table}[h!]
\centering
\begin{tabular}{|lll|}
\hline
Tool Name & Submitter(s) & Core Technology \\
\hline
 & &\\
\multicolumn{3}{|c|}{\textbf{\emph{Preprocessors (4)}}} \\
 & &\\
\preproA & A.\ Van Gelder &failed literal detection\\
\preproC & A.\ Van Gelder &univ. expansion, variable elimination\\
\preproB & M.\ Seidl, A.\ Biere & univ. expansion, variable elimination  \\
\preproD & M.\ Narizzano & variable elimination, equivalence rewriting\\
\hline
 & &\\
\multicolumn{3}{|c|}{\textbf{\emph{Solvers (15)}}} \\
 & & \\
\includegraphics[scale=0.8]{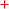} \sqube & M. Narizzano & CDCL\\
\includegraphics[scale=0.8]{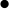} \ghost &  W. Klieber & dual prop.\\
\includegraphics[scale=0.8]{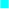} \ghostc & W. Klieber & dual prop.~and CEGAR\\
\includegraphics[scale=0.8]{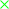} \bghostc & W. Klieber &dual prop., abstraction refinement, \bloqqer\\
\includegraphics[scale=0.8]{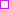} \rareqs & M. Janota & expansion and abstraction refinement (CEGAR)\\
\includegraphics[scale=0.8]{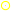} \hiqqer & A. Van Gelder & preprocessing and CDCL\\
\includegraphics[scale=0.8]{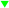} \ooqm & A. Goultiaeva & CDCL\\
\includegraphics[scale=0.8]{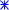} \dooqm & A. Goultiaeva & struct. rec., CDCL, dual prop.\\
\includegraphics[scale=0.8]{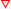} \sdooq & A. Goultiaeva & struct. rec, CDCL, dual prop., preprocessing\\
\includegraphics[scale=0.8]{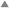} \nenofexm & F. Lonsing & existential expansion \\
\includegraphics[scale=0.8]{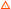} \depqbfm & F. Lonsing & CDCL\\
\includegraphics[scale=0.8]{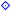} \ldepqbf & F. Lonsing & lazy CDCL\\
\freetoqbf (2QBF)* & S. Bayless &augmented SAT solver, special decision heuristic\\
\minitoqbf (2QBF)* & S. Bayless &augmented SAT solver\\
\minitoqbf ext.~(2QBF)* & S. Bayless & augmented SAT solver, preprocessing\\
 & &\\
\hline 
& &\\
\multicolumn{3}{|c|}{\textbf{\emph{Certification Tools (2)}}} \\
& &\\
\qbfcert & A. Niemetz, & extraction from resolution proofs\\
& M. Preiner & \\
\resqu & V. Balabanov, &  extraction from resolution proofs\\
& J.-H. R. Jiang &\\
&& \\
* track skipped & &\\
\hline

\hline
\end{tabular}
\caption{Tools and Solvers participating in the QBF Gallery 2013.}
\label{tab:tools}
\end{table}

\subsubsection{Tools} \label{sec:tools}

An overview of the submitted tools and contributors 
is shown in Table~\ref{tab:tools}. The submissions include 
four preprocessors, 15 solvers as well as two certification tools. 
Please note that it was allowed to submit up to 
three different configurations of one tool.

\paragraph{Preprocessors} 
The goal of preprocessors is to rewrite a formula 
in prenex conjunctive normal form
such that
(1) its truth value is not changed and (2) it becomes easier to solve.
To this end, preprocessors try to remove irrelevant information 
and to enhance the formula with additional structure useful 
for the solving process. Therefore, preprocessing 
might not only modify and eliminate clauses of a formula, but also 
add new clauses and even introduce new variables.
Four preprocessors were submitted to the QBF Gallery: 
\bloqqer\footnote{\url{http://fmv.jku.at/bloqqer/}},  
\hiqqerp, \hiqqere, and \squeezebf. 
They all implement standard optimization techniques
like pure and unit literal detection, universal reduction as 
well as equivalence substitution. \hiqqerp is a tuned version of 
\bloqqer~\cite{DBLP:conf/cade/BiereLS11} 
that implements variable elimination, universal expansion
and blocked clause elimination amongst other techniques. 
\squeezebf~\cite{DBLP:conf/sat/GiunchigliaMN10} also uses
variable elimination and additionally some  special kind of 
equivalence rewriting that recovers structure lost during 
the normal form transformation. 
\hiqqere~\cite{DBLP:conf/sat/GelderWL12} uses an extension of 
failed literal detection.

\paragraph{Solvers} 
Table~\ref{tab:tools} provides an overview of the submitted solvers. 
The 
icons shown are later used in the plots to indicate the performance of a solver.
Previously, QBF competitions had a CNF track, 
a non-CNF track as well as a 2-QBF track. We also planned to organize
these three different tracks, but due to the lack of submissions 
in the non-CNF track and the 2-QBF track, we focused on 
CNF solvers.  Solver developers were allowed to submit three 
variants or configurations of each solver. 
Four contributors exercised this option, 
which includes versions of solvers that were enhanced by third-party 
preprocessors as well.  The solver
\sqube was the current version of QuBE~\cite{DBLP:journals/jsat/GiunchigliaMN10}, one of the dominators of the former QBF competitions.   
The preprocessor \squeezebf is part of \sqube, where \squeezebf was also
submitted as a standalone tool 
(see above). The solver \sqube is based on the clause/cube learning (CDCL)
variant for QBF. The solver \depqbf~\cite{LonsingPhD2012} also implements clause/cube 
learning and additionally it considers variable independencies reconstructed
from the formula structure to gain more flexibility during the solving process. 
The variant  \ldepqbf~\cite{qpupacceptedsat13} uses a different learning approach.
The solver \ooqm~\cite{DBLP:conf/sat/GoultiaevaB13} also implements clause and cube learning. Additionally, 
 \dooqm implements dual propagation~\cite{DBLP:conf/date/GoultiaevaSB13}
by reconstructing structural information from the CNF. Finally, 
\sdooq uses the preprocessor \squeezebf before solving.
The solver \hiqqer combines the preprocessor \bloqqer with failed literal 
detection~\cite{DBLP:conf/sat/GelderWL12}. If the formula is not solved by preprocessing, then 
an adopted version of the complete solver \depqbf is called.  
The \ghost solver~\cite{ghost} aims at
overcoming the loss of structural information 
imposed by the transformation to PCNF by introducing a concept called 
``ghost variables''. These ghost variables may be considered as 
a dual variant of the Tseitin variables and provide an efficient 
mechanism to simulate reasoning on disjunctive normal form.
The solver \ghostc~\cite{ghost2} extends \ghost with an additional
learning technique based on
counterexample-guided abstraction-refinement (CEGAR).
The variant \bghostc calls the preprocessor \bloqqer in certain situations.
The solver \rareqs applies CEGAR in an expansion-based approach~\cite{ghost2}.

\paragraph{Certification Frameworks}

\qbfcert~\cite{DBLP:conf/sat/NiemetzPLSB12} and
\resqu~\cite{DBLP:journals/fmsd/BalabanovJ12} are tool suites to produce
Skolem-function models of satisfiable QBFs and Herbrand-function countermodels
of unsatisfiable QBFs. To this end, these tools extract a (counter)model from
a resolution proof of (un)satisfiability. Since \qbfcert and \resqu were the
only certification tools submitted, we decided to consider additional publicly
available tools and to run additional experiments as presented in
Section~\ref{sec:cert}.

\begin{table}[t]
\centering
\begin{tabular} {|l|l||r|rrr|rr|}
\hline
name & showcase & \# & vars$^*$ & clauses$^*$ & alt$^*$ & $\exists^*$ & $\forall^*$\\
\hline
\hline
\texttt{bomb} & applications &150&3234&210265&3&3220&14\\
\texttt{dungeon} & applications & 150&36324&264222&3&36318&5\\
\texttt{reduction-finding} & applications & 150&1777&8191&2&1731&44 \\
\texttt{planning-CTE} & applications & 150&3239&600112&5&3237&2 \\
\texttt{qbf-hardness} & applications & 150&2299&8457&22&2058&100 \\
\texttt{sauer-reimer} & applications & 150&13655&40092&3&13407&248\\
\hline
\texttt{AABBCCDD} & preprocessing & 234&12060&44516&6&6464&845 \\
\texttt{AADDBBCC} & preprocessing & 241&12409&45522&6&6353&816 \\
\hline
\texttt{eval2012r2} & solving & 345&32924&77709&14&20414&733 \\
\texttt{eval2012r2} with \bloqqer & solving & 276&6834&34938&6&6077&756\\
\texttt{eval2010} & solving & 568&23546&53857&43&18337&223\\
\texttt{eval2010} with \bloqqer & solving & 420&3532&22578&9&3329&203\\
\hline
\end{tabular}
\caption{Formula characteristics of the different benchmark sets: 
number of formulas (\#), average number of variables (vars), 
average number of clauses (clauses), average number of quantifier alternations
(alt), average number of universal/existential variables ($\exists/\forall$). }
\label{tab:formulas}
\end{table}

\subsubsection{Benchmarks} \label{sec:benchmarks}

In the following, we describe the benchmark sets used in
our experiments. New benchmark sets submitted by the participants as well as 
 benchmarks from the public QBFLIB repository\footnote{\url{http://www.qbflib.org/}} were considered. 
In our experiments, all formulas are in prenex conjunctive 
normal form (PCNF) with a quantifier prefix having an arbitrary number 
of quantifier alternations. Details on syntactic formula characteristics are 
shown in Table~\ref{tab:formulas}. For the showcase on applications we selected 
benchmark sets consisting of 150 formulas each.

\begin{itemize}

\item{Set \texttt{eval2010}:} the complete set of 568 formulas used for 
\emph{QBFEVAL 2010}~\cite{DBLP:conf/sat/PeschieraPTBKL10}.

\item{Set \texttt{eval2012r2}:} 345 formulas sampled from the collection of formulas available from
QBFLIB. This set was also used for the QBF 
competition \emph{QBFEVAL 2012 Second Round}, 
an unofficial repetition of the QBFEVAL 2012
with a new benchmark set.\footnote{\url{http://fmv.jku.at/seidl/qbfeval2012r2/}} 

\item{Set \texttt{eval2012r2-inc-preprocessed}:} Instances from the 
set \texttt{eval2012r2} which were obtained by repeated, incremental
preprocessing using the four preprocessors that were submitted to the showcase
on preprocessing, as described in Section~\ref{sec:prepro}. 
We obtained the following two sets:

\begin{itemize}
\item{Set \texttt{AABBCCDD}:} 234 instances
resulting from the set \texttt{eval2012r2} by incremental preprocessing, where the preprocessors
are called in a tool chain in at most six rounds. From
the 345 instances, 111 instances were solved during incremental preprocessing. In the tool chain, the formula produced by one preprocessor is
forwarded to the next. A wall-clock time limit of 120
seconds was set for each call of a preprocessor. The preprocessors were
executed in the ordering AABBCCDD, where ``A'' is \preproA, ``B'' is \preproB, 
``C'' is \preproC, and ``D'' is \preproD. Fixpoint detection was implemented so
that preprocessing stops if the formula is no longer modified by any 
preprocessor.

\item{Set \texttt{AADDBBCC}:} 241 instances
resulting from the set \texttt{eval2012r2} by incremental preprocessing. From
the 345 instances, 104 instances were solved during 
preprocessing. This set was generated in similar fashion as the set
\texttt{AABBCCDD} except that the execution
ordering AADDBBCC was used.
\end{itemize}

We selected the execution orderings \texttt{AABBCCDD} and \texttt{AADDBBCC}
based on empirical findings we made in the showcase on preprocessing. For
example, with ordering \texttt{AABBCCDD} the largest number of instances was
solved. Due to the different characteristics of the techniques implemented in
\preproB (``B'') and \preproD (``D'') we selected \texttt{AADDBBCC}, where
\preproD is executed before \preproB.

\item{Set \texttt{reduction-finding}:} formulas
generated from instances of reduction finding~\cite{DBLP:conf/birthday/CrouchIM10,qbf2013/JordanKaiser,DBLP:conf/sat/JordanK13},
which is the problem to determine whether parametrized quantifier-free
reductions exist between various decision problems in NL for one set of fixed
parameters. A program to generate this set of benchmarks was also submitted by
Charles Jordan and Lukasz Kaiser. 
 The submitted set consists of $4608$ QBF encodings of $2304$ reduction
problems, where each problem is encoded as a QBF with prefix
$\forall\exists$. 

\item{Set \texttt{conformant-planning}:} 1750 instances from a planning domain
with uncertainty in the initial state, contributed by Martin Kronegger,
Andreas Pfandler, and Reinhard Pichler~\cite{qbf2013/KroneggerPfandler}.
The consists of two different kinds of planning problems: \texttt{dungeon}
and \texttt{bomb}.

\item{Set \texttt{planning-CTE}:} 150 instances resulting from 
  compact tree encodings (CTE) of planning problems, contributed by Michael Cashmore~\cite{DBLP:conf/ecai/CashmoreFG12}.

\item{Set \texttt{sauer-reimer}:} 924 instances from QBF-based test generation, contributed by Paolo
Marin~\cite{DBLP:conf/aspdac/SauerRPSB13}.

\item{Set \texttt{qbf-hardness}:} 198 instances from bounded model checking of incomplete designs,
contributed by Paolo Marin~\cite{DBLP:conf/sat/MillerKLB10}.

\item{Set \texttt{samples-eval12r2}: ten sets containing 461 formulas
    each. The formulas in these sets were randomly selected from the instances
    available in QBFLIB. The random selection process was carried out with
    respect to families of instances, thus avoiding that instances from
    large families are overrepresented in the sampled set.
In general it can often be observed that solvers perform either very good or 
very bad on a specific family. Therefore, data accumulated based on
benchmark sets which are biased towards particular families is not
expressive and may lead to misleading conclusions.

}
\end{itemize}

%%%%%%%%%%%%%%%%%%%%%%%%%%%%%%%%%%%%%%%%%%%%%%%%%%%%%%%%%%%%%%%%%%%%%%%%%%%%%%%%
%%%%%%%%%%%%%%%%%%%%%%%%%%%%%%%%%%%%%%%%%%%%%%%%%%%%%%%%%%%%%%%%%%%%%%%%%%%%%%%%
%\newpage
\section{Showcases in the QBF Gallery 2013}

We invited the participants to suggest 
the showcases to be considered. At the end, four different 
showcases were considered: (1) solving, (2) preprocessing, (3)
applications, and (4) certification. In the following, 
we outline the setup of each showcase
and summarize the most important results.

\subsection{Showcase: Solving}
\label{sec:solving}

In this showcase, we evaluated the solvers on the benchmark set 
 \texttt{eval2012r2}. 
 We carried out separate runs with and without preprocessing using 
the preprocessor \bloqqer. All experiments were run on a 
64-bit Linux Ubuntu 12.04 system with an Intel
Core 2 Quad Q9550@2.83GHz and 8GB of memory.
We used a time limit of $900$ seconds and a memory
limit of $7$ GB. In the following, we focus on the results obtained for the set
\texttt{eval2012r2}. Additional results for the set \texttt{eval2010},
including tables and plots, are reported in \ref{sec:appendix:solving}.
While some formulas appear in both sets \texttt{eval2010} and \texttt{eval2012r2}, \texttt{eval2012r2} contains 
formulas from more recent benchmark sets. 

\begin{figure}[t!]
\centering
\includegraphics{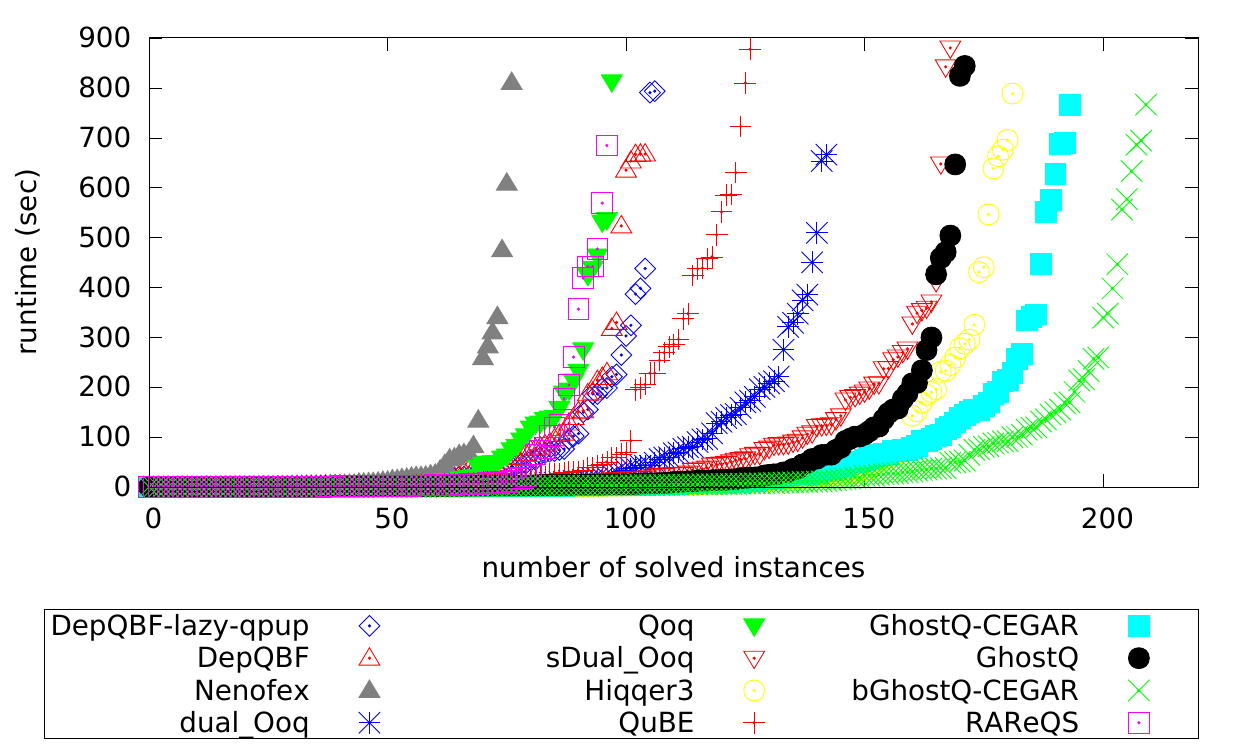}
\caption{Sorted runtimes of the solvers on the
  benchmark set \texttt{eval2012r2} without prior preprocessing using \bloqqer
  (related to Table~\ref{eval12}).}
\label{cactuseval12plot}
\end{figure}
\begin{table}[h!]
\centering
 \begin{tabular}{|l|r|r|r|r|r|r|}
\hline
&\multicolumn{4}{c|}{\textbf{number of solved formulas}} & \multicolumn{2}{c|}{\textbf{runtime (sec)}}\\\hline
\textbf{solver}&solved&sat&unsat&unique&avg & total\\
\hline
\bghostc&210&111&99&0&50&132K\\
\ghostc&194&103&91&0&55&146K\\
\hiqqer&182&93&89&6&51&156K\\
\ghost&172&87&85&2&50&164K\\
\sdooq&169&80&89&5&63&169K\\
\dooqm&143&66&77&0&58&190K\\
\sqube&127&60&67&0&93&207K\\
\ldepqbf&107&43&64&0&63&220K\\
\depqbfm&105&42&63&0&73&223K\\
\ooqm&98&34&64&0&63&228K\\
\rareqs&97&34&63&5&97&228K\\
\nenofexm&77&34&43&5&53&245K\\
\hline
 \end{tabular}
 \caption{Solving statistics for the set \texttt{eval2012r2} (345 instances)
   without prior preprocessing using \bloqqer. Some solvers like \dooqm and
   \hiqqer, for example,
   apply built-in preprocessing. The table shows the total number of solved
   instances (column ``solved''), solved satisfiable (column ``sat'') and
   unsatisfiable ones (column ``unsat''),
    uniquely
 solved instances (column ``unique''), and average runtime of solved formulas  and total runtime (columns
 ``avg'' and ``total'') on the whole benchmark set. }
\label{eval12}
\end{table}

Table~\ref{eval12} shows 
detailed results for the set \texttt{eval2012r2} 
without preprocessing.
Note that some solvers like \dooqm and \hiqqer, for example, apply built-in
preprocessing. Columns ``runtime'' report the average runtime of solved
formulas and the total runtime spent on the entire benchmark set.

This benchmark set is very suitable for  search-based solvers such as
\bghostc (and its variants) and \hiqqer, while the performance of expansion-based solvers like
\rareqs and \nenofexm is worse. However, both \rareqs and \nenofexm
 solved five unique instances that no other solver could
 solve. Table~\ref{eval12suite} presents detailed numbers of solved instances for each
 benchmark family in the set \texttt{eval2012r2}. Figure~\ref{cactuseval12plot} shows a
 cactus plot of the runtimes of the solvers.

\begin{figure}[h!]
\centering
\includegraphics{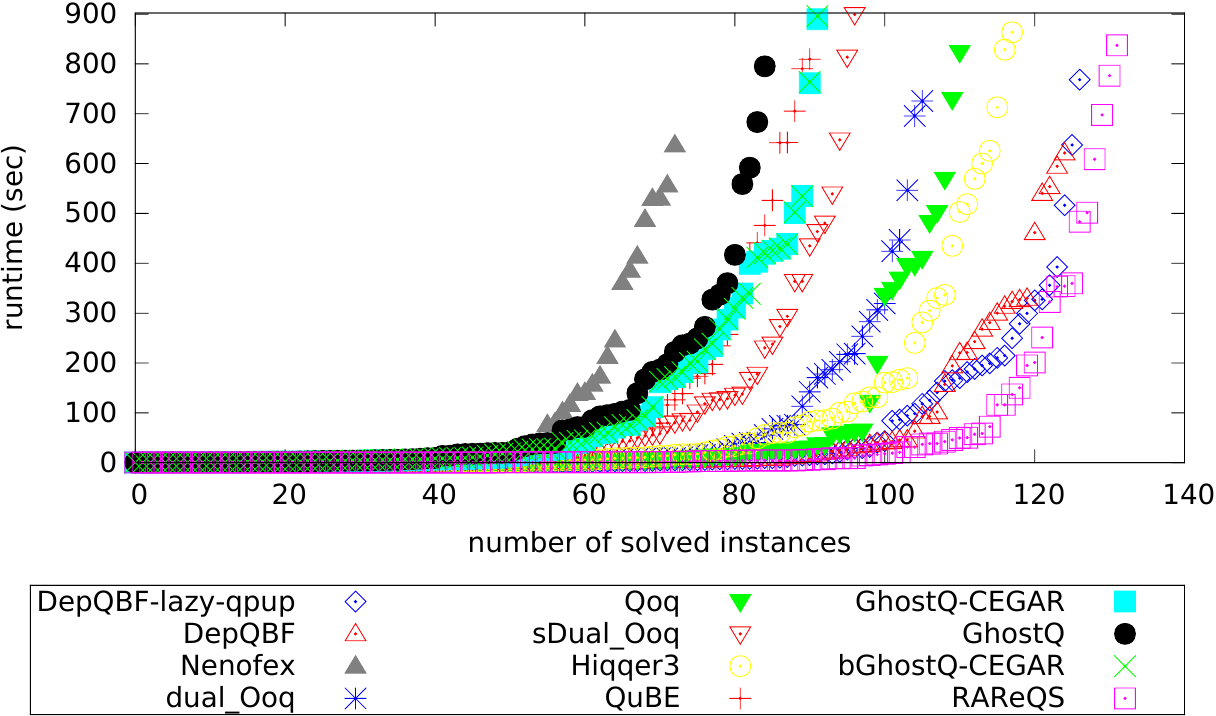}
\caption{Sorted runtimes of the solvers on the
  benchmark set \texttt{eval2012r2} with prior preprocessing using \bloqqer (related to Table~\ref{eval12bloqqerpp}).}
\label{cactuseval12bloqqerppplot}
\end{figure}

We obtain a very different picture of the solver performance when the set
\texttt{eval2012r2} is preprocessed using \bloqqer. In the following
experiment, every solver is run on the $276$ instances that were preprocessed
but not solved by \bloqqer. Some solvers additionally apply their built-in
preprocessors. Table~\ref{eval12bloqqerpp} shows the number of successfully solved formulas 
which, compared to Table~\ref{eval12}, gives a very different picture. 
Notably, \rareqs and
\ldepqbf (and its variants) are now more highly ranked than \bghostc (and its
variants). Although \nenofexm still solved the smallest number of instances,
as in Table~\ref{eval12}, it solved eight instances uniquely, which is the
largest number of uniquely solved instances among all
solvers. Table~\ref{eval12_bloqqer_ppsuite} and
Figure~\ref{cactuseval12bloqqerppplot} show detailed, family-based statistics
and runtimes, respectively.

\begin{table}[ht]
\centering
\small
 \begin{tabular}{|l|r|r|r|r|r|r|r|r|r|r|r|r|}
\hline

&
\begin{sideways}\sqube\end{sideways}&
\begin{sideways}\bghostc\end{sideways}&
\begin{sideways}\dooqm\end{sideways}&
\begin{sideways}\rareqs\end{sideways}&
\begin{sideways}\ghostc\end{sideways}&
\begin{sideways}\hiqqer\end{sideways}&
\begin{sideways}\ghost\end{sideways}&
\begin{sideways}\depqbfm\end{sideways}&
\begin{sideways}\nenofexm\end{sideways}&
\begin{sideways}\sdooq\end{sideways}&
\begin{sideways}\ooqm\end{sideways}&
\begin{sideways}\ldepqbf\end{sideways}\\
\hline

Ansotegui (20)
&11&5&6&6&5&10&7&10&0&11&10&10\\

Ayari (12)
&6&5&6&5&0&7&0&0&6&6&4&0\\

Basler (18)
&3&13&1&0&13&6&4&0&0&8&0&0\\

Biere (16)
&5&14&9&1&14&5&14&3&3&6&3&3\\

gelder (12)
&9&12&11&6&12&12&12&11&3&11&10&11\\

Gent-Rowley (4)
&1&1&1&0&1&1&1&1&0&1&1&1\\

Herbstritt (12)
&10&9&10&7&9&9&10&6&1&10&6&6\\

jiang (12)
&2&5&6&1&5&6&5&5&1&6&6&4\\

Katz (12)
&2&0&0&0&0&2&0&0&1&3&0&0\\

Kontchakov (18)
&12&8&16&0&8&18&0&18&0&17&1&18\\

Lahiri-Seshia (3)
&0&2&0&0&2&0&2&0&0&0&0&0\\

Letombe (6)
&6&6&6&6&6&6&6&6&2&6&6&6\\

Ling (2)
&0&2&2&2&2&2&1&2&2&2&2&2\\

Mangassarian-Veneris (11)
&2&4&4&6&4&6&3&5&7&4&4&6\\

Messinger (6)
&0&0&0&0&0&0&0&0&0&0&0&0\\

Miller-Marin (18)
&14&15&12&17&15&17&13&15&12&16&12&16\\

Mneimneh-Sakallah (38)
&15&31&18&0&31&9&31&0&0&16&0&0\\

Palacios (16)
&2&9&5&14&9&5&4&5&9&5&5&4\\

Pan (48)
&10&33&19&6&30&39&30&8&15&22&18&8\\

Rintanen (12)
&2&8&6&10&8&6&8&7&10&5&6&9\\

sauer$\_$reimer (6)
&2&6&2&2&6&4&5&1&0&2&2&1\\

Scholl-Becker (25)
&2&13&3&8&13&3&15&2&4&4&2&2\\

Wintersteiger (18)
&11&9&0&0&1&9&1&0&1&8&0&0\\
\hline \hline

total (345)
&
127&
210&
143&
97&
194&
182&
172&
105&
77&
169&
98&
107\\
\hline
 \end{tabular}
 \caption{Numbers of solved instances for each benchmark family in the set
   \texttt{eval2012r2} (related to the results reported in
   Table~\ref{eval12}). The number of instances in each family (first column) is shown in
   parentheses.}
\label{eval12suite}
\end{table}

%\begin{center}

\begin{table}[h!]
\centering
 \begin{tabular}{|l|r|r|r|r|r|r|}
\hline
&\multicolumn{4}{c|}{\textbf{number of solved formulas}} & \multicolumn{2}{c|}{\textbf{runtime (sec)}}\\\hline
\textbf{solver}&solved&sat&unsat&unique&avg&total\\
\hline
\rareqs&132&66&66&7&55&136K\\
\ldepqbf&127&66&61&0&55&141K\\
\depqbfm&125&66&59&0&55&142K\\
\hiqqer&118&59&59&3&82&151K\\
\ooqm&111&58&53&3&58&155K\\
\dooqm&106&57&49&2&58&159K\\
\sdooq&97&54&43&0&86&169K\\
\ghostc&92&54&38&0&99&169K\\
\bghostc&92&54&38&0&99&174K\\
\sqube&91&52&39&0&96&175K\\
\ghost&85&50&35&0&88&179K\\
\nenofexm&73&39&34&8&77&188K\\
\hline
 \end{tabular}
 \caption{Solving statistics for the set \texttt{eval2012r2} preprocessed with
   \bloqqer. After preprocessing, $276$ instances remained unsolved. The ranking of the solvers largely differs from the ranking shown
   in Table~\ref{eval12} where preprocessing prior to solving was omitted.}
\label{eval12bloqqerpp}
\end{table}

\begin{table}[ht]
\centering
\small
 \begin{tabular}{|l|r|r|r|r|r|r|r|r|r|r|r|r|}
\hline

&
\begin{sideways}\sqube\end{sideways}&
\begin{sideways}\bghostc\end{sideways}&
\begin{sideways}\dooqm\end{sideways}&
\begin{sideways}\rareqs\end{sideways}&
\begin{sideways}\ghostc\end{sideways}&
\begin{sideways}\hiqqer\end{sideways}&
\begin{sideways}\ghost\end{sideways}&
\begin{sideways}\depqbfm\end{sideways}&
\begin{sideways}\nenofexm\end{sideways}&
\begin{sideways}\sdooq\end{sideways}&
\begin{sideways}\ooqm\end{sideways}&
\begin{sideways}\ldepqbf\end{sideways}\\
\hline

Ansotegui (20)
&12&9&6&7&9&14&11&12&2&7&13&12\\

Ayari (6)
&0&0&0&0&0&1&0&0&4&0&0&0\\

Basler (12)
&0&0&0&3&0&1&0&0&0&0&5&0\\

Biere (14)
&2&3&3&5&3&3&3&3&3&3&4&4\\

gelder (6)
&6&6&6&6&6&6&6&6&6&6&6&6\\

Gent-Rowley (4)
&1&1&1&0&1&1&1&1&0&1&1&1\\

Herbstritt (11)
&7&0&7&11&0&8&0&9&0&6&8&9\\

jiang (12)
&3&6&6&5&6&6&6&4&2&6&6&6\\

Katz (12)
&1&1&3&0&1&2&2&3&1&3&2&3\\

Kontchakov (18)
&9&2&2&1&2&12&3&18&0&1&1&18\\

Lahiri-Seshia (3)
&0&0&0&0&0&0&0&0&0&0&0&0\\

Letombe (6)
&5&6&6&6&6&6&6&6&1&6&6&6\\

Ling (2)
&2&2&2&2&2&2&2&2&2&2&2&2\\

Mangassarian-Veneris (10)
&3&2&4&5&2&5&3&5&9&4&4&5\\

Messinger (6)
&0&0&0&0&0&0&0&0&0&0&0&0\\

Miller-Marin (11)
&8&11&9&11&11&10&9&9&7&9&9&9\\

Mneimneh-Sakallah (37)
&14&14&30&26&14&9&6&16&3&24&19&15\\

Palacios (16)
&2&5&5&11&5&5&5&5&10&5&5&5\\

Pan (20)
&11&6&6&11&6&12&7&10&7&6&7&9\\

Rintanen (12)
&2&9&4&10&9&8&7&8&10&3&7&9\\

sauer$\_$reimer (4)
&2&3&2&2&3&2&3&2&0&2&2&2\\

Scholl-Becker (25)
&1&6&4&9&6&5&5&6&6&3&4&6\\

Wintersteiger (9)
&0&0&0&1&0&0&0&0&0&0&0&0\\
\hline \hline
total (276)
&
91&
92&
106&
132&
92&
118&
85&
125&
73&
97&
111&
127\\
\hline
 \end{tabular}
 \caption{Numbers of solved instances for each benchmark family in the set
   \texttt{eval2012r2} preprocessed with \bloqqer (related to the results
   reported in Table~\ref{eval12bloqqerpp}). The number of instances in each family (first column) is shown in
   parentheses.}
\label{eval12_bloqqer_ppsuite}
\end{table}

As illustrated by the different rankings in Tables~\ref{eval12}
and~\ref{eval12bloqqerpp}, the performance of the solvers varies depending on
the use of preprocessing. 
On the original set \texttt{eval2012r2}  solvers with built-in preprocessing 
outperform solvers that do not have built-in preprocessing, as shown in
Table~\ref{eval12}. On the preprocessed set \texttt{eval2012r2}, another
built-in preprocessing step using \bloqqer might have little effect, cause
runtime overhead, and hence harm the performance of a solver. Solvers like \dooqm and \ghost try to reconstruct part of the structure
of a CNF formula as a preprocessing step. The purpose of this reconstruction step
is to recover the structure that was obscured during the conversion of a
non-CNF formula into CNF. Structural information might improve the performance
of CNF-based
solvers~\cite{DBLP:conf/sat/GoultiaevaB13,DBLP:conf/date/GoultiaevaSB13}. However,
structure reconstruction might be hindered when preprocessing by \bloqqer is
applied upfront as in the experiment shown in Table~\ref{eval12bloqqerpp}.
The solver \bghostc dedicates some of its solving time to 
run \bloqqer in order to see if \bloqqer is able to solve 
a formula quickly. If this is not the case, then the original 
formula is considered for solving and the work spent for the 
preprocessing was useless.

%\end{center}

%\clearpage

\subsubsection{Preprocessing and Solving: ``Best Foot Forward'' Analysis}
\label{sec:best:worst:foot}

On the one hand, solvers that do not apply built-in preprocessing might
perform better on an instance that has been preprocessed using \bloqqer. On
the other hand, solvers with built-in preprocessing or structure
reconstruction might prefer the original instance.

In order to analyze the performance of the solvers with and without prior
preprocessing in more detail, we carried out the following experiments. We
ran \bloqqer on all 345 instances in the benchmark set
\texttt{eval2012r2}. From all these instances, 276 remained unsolved by
\bloqqer. We ran each solver twice: once on the 276 instances after they have
been preprocessed by \bloqqer, and once on the respective 276 original
instances from the set \texttt{eval2012r2} without preprocessing. That is, in
this experiment we exclude instances from the set \texttt{eval2012r2} that
were solved already by \bloqqer. 

We classified the solvers into two categories, depending on the numbers of
instances that were solved in these two runs. We classified a solver in the
``NO \bloqqer'' category if it performs better on the original instances than
on the instances that have been preprocessed. If a solver performs better on
the preprocessed instances than on the original ones, then we classified it in
the ``WANT \bloqqer'' category.

\begin{table}[t]
\begin{center}
\begin{tabular}{lrr}
\hline
       & \multicolumn{2}{c}{\emph{Number Solved}}\\
Category/& Best& Worst\\
Solvers& Foot& Foot\\
\hline
\emph{NO \bloqqer (solvers perform better without \bloqqer)} & & \\
 \bghostc & 142& 93\\
 \ghostc & 142& 93\\
 \ghost & 122& 84\\
 \sdooq & 118& 99\\
  \dooqm & 105& 89\\

\hline

\emph{WANT \bloqqer (solvers perform better with \bloqqer)} & & \\
 \rareqs & 132& 79\\
 \ldepqbf & 128& 88\\
 \depqbfm & 125& 86\\
 \hiqqer & 117& 113\\
 \ooqm & 93& 65\\
 \sqube & 91& 90\\
 \nenofexm & 68& 50\\
\hline
\end{tabular}
\end{center}
\caption{Classification of solvers into two categories depending on their
performance on 276 instances of the set \texttt{eval2012r2} with (category
``WANT \bloqqer'') and without prior preprocessing by \bloqqer (category ``NO
\bloqqer''). In each of these categories, column ``Best Foot'' shows the
numbers of instances that were solved when choosing to run on preprocessed
instances or on original ones, respectively. Column ``Worst Foot'', on the
contrary, shows the numbers of instances solved when making the opposite choice.
}
\label{bestfooteval2012r2RECOMPUTED}
\end{table}

Table~\ref{bestfooteval2012r2RECOMPUTED} shows the final classification of the
solvers. In the category ``WANT \bloqqer'', the columns ``Best Foot'' and
``Worst Foot'' report the numbers of instances that were solved by a solver
\emph{with and without} prior preprocessing by \bloqqer, respectively. In
contrast to that, in the category ``NO \bloqqer'', these columns report the
numbers of instances that were solved by a solver \emph{without and with}
prior preprocessing by \bloqqer, respectively. That is, column ``Best Foot''
represents the best choice of a solver in terms of solved instances whether to
run on the original instances or on the preprocessed ones. Contrary to
the best choice, column ``Worst Foot'' represents the respective worst choice in
each category.

It is interesting to note that \hiqqer and \sqube do not have much preference
whether to run on original instances or on preprocessed ones, because their
respective best foot and worst foot statistics differ only by four and one
formula(s).
Recall that \hiqqer includes a modified variant of \bloqqer and that
\sqube applies a powerful built-in preprocessor. We obtained different results  
when considering a different benchmark set (cf.\ Table~\ref{bestfooteval2010RECOMPUTED} in the appendix).

The classification in Table~\ref{bestfooteval2012r2RECOMPUTED} confirms the
trend that is indicated by the different rankings of the solvers in
Tables~\ref{eval12} and~\ref{eval12bloqqerpp}. Solvers in the ``NO \bloqqer''
category like \bghostc (and its variants) perform better without prior
preprocessing and thus are more highly ranked in Table~\ref{eval12} than
solvers in the ``WANT \bloqqer'' category like \rareqs and \ldepqbf. In
contrast to that, solvers in the ``WANT \bloqqer'' category perform better
with prior preprocessing and thus dominate the solvers from the ``NO
\bloqqer'' category in the ranking shown in Table~\ref{eval12bloqqerpp}.

The best-foot-forward analysis presented above revealed that the performance of the
solvers might heavily depend on the use of preprocessing when applied before
the actual solving. In the related experiments, we preprocessed the instances
by a single application of \bloqqer. We could  also combine several
preprocessing tools like \squeezebf, for example, with \bloqqer to analyze
their combined effects. However, here we decided to focus on \bloqqer since some of
the submitted solvers like \bghostc already apply \bloqqer as a built-in
preprocessor. In the showcase on preprocessing to be presented below
(Section~\ref{sec:prepro}), we report on a comprehensive evaluation of several
preprocessing tools both independently from solvers and combinations thereof.

\begin{table}[t!]
\begin{center}
{\scriptsize
\begin{tabular}{|lr|lr|lr|lr|lr|lr|lr|}
\hline
\multicolumn{2}{|c|}{\bf Trial 1}&
\multicolumn{2}{c|}{\bf Trial 2}&
\multicolumn{2}{c|}{\bf Trial 3}&
\multicolumn{2}{c|}{\bf Trial 4}&
\multicolumn{2}{c|}{\bf Trial 5}&
\multicolumn{2}{c|}{\bf Trial 6}&
\multicolumn{2}{c|}{\bf Trial 7}
\\
\hline
BB& 312& BB& 317& BB& 316& BB& 315& BB& 315& BB& 310& BB& 319\\
HH& 293& HH& 298& HH& 300& HH& 298& HH& 291& HH& 290& HH& 299\\
AA& 284& AA& 285& AA& 274& AA& 272& AA& 279& AA& 275& AA& 284\\
CC& 274& CC& 270& CC& 265& CC& 270& CC& 270& CC& 273& CC& 274\\
DD& 241& DD& 237& DD& 236& DD& 242& DD& 232& DD& 241& DD& 239\\
EE& 211& EE& 211& EE& 205& EE& 216& EE& 208& EE& 215& EE& 211\\
GG& 204& GG& 201& GG& 200& GG& 202& GG& 194& GG& 194& GG& 207\\
FF& 159& FF& 162& FF& 161& FF& 162& FF& 159& FF& 156& FF& 171\\
\hline
\end{tabular}
}
\end{center}
\caption{Ranking of solvers by numbers of instances solved in seven randomly
  sampled benchmark sets (trials), each containing 455 instances. Solver names are
  anonymized using a two-letter code.}
\label{strat:sampling:solved}
\end{table}

\begin{table}[t!]
\begin{center}
{\scriptsize
\begin{tabular}{|lr|lr|lr|lr|lr|lr|lr|}
\hline
\multicolumn{2}{|c|}{\bf Trial 1}&
\multicolumn{2}{c|}{\bf Trial 2}&
\multicolumn{2}{c|}{\bf Trial 3}&
\multicolumn{2}{c|}{\bf Trial 4}&
\multicolumn{2}{c|}{\bf Trial 5}&
\multicolumn{2}{c|}{\bf Trial 6}&
\multicolumn{2}{c|}{\bf Trial 7}
\\
\hline
BB& 637	&BB&	614	&BB&	619	&BB&	625&	BB&	622	&BB	&646	&BB&	605	\\
HH&	723	&HH&	698	&HH&	693	&HH&	701	&HH&	729	&HH&	734	&HH&	695	\\
AA&	759	&AA&	753	&AA&	802	&AA&	810	&AA&	781	&AA&	796	&AA&	757	\\
CC&	798	&CC&	815	&CC&	838	&CC&	816	&CC&	815	&CC&	804	&CC&	799	\\
DD&	947	&DD&	963	&DD&	969	&DD&	943	&DD&	986	&DD&	947	&DD&	955	\\
EE&	1077	&EE&	1077	&EE&	1104	&EE&	1055	&EE&	1090	&EE&	1060	&EE&	1078	\\
GG&	1109	&GG&	1122	&GG&	1128	&GG&	1119	&GG&	1154	&GG&	1153	&GG&	1096	\\
FF&	1304	&FF&	1292	&FF&	1296	&FF&	1291	&FF&	1305	&FF&	1318	&FF&	1252	\\

\hline
\end{tabular}
}
\end{center}
\caption{Like Table~\ref{strat:sampling:solved}, but solvers are ranked
  according to their penalized average runtime (PAR10).}
\label{strat:sampling:times}
\end{table}

\subsubsection{Stratified Sampling}

In addition to the application of preprocessing prior to solving, the actual
selection of benchmarks might have an influence on the performance of solvers
and hence on the rankings in terms of solved instances. A ranking of solvers
obtained by experiments might be skewed if certain families of instances are
overrepresented in the benchmark set that underlies the experimental
evaluation. In order to analyze the effect of the benchmark selection on the
ranking of solvers, we carried out the following sampling experiment.

From all benchmark  instances available at QBFLIB, we randomly sampled
seven benchmark sets containing 455 instances each in a stratified manner. The
stratification consisted of randomly choosing six instances from each 
benchmark family. Thereby, as noted
above, we consider a set of instances a family if this set was classified as
such at the time the set was submitted to QBFLIB. The instances in the sampled
sets were as contributed by users for use as benchmarks, and we did not apply
preprocessing by \bloqqer.

In this experiment, we deliberately do not disclose the actual names of the
solvers, but used two-letter names. The intention was to put the focus 
on the experiment itself (which is the evaluation of the benchmark set) and
not on the evaluation of a particular solving technique. 
Based on the best-foot-forward experiment described above, we 
selected the eight solvers that solved the largest number of formulas. 
Our focus was on understanding the effects that different selections
of benchmark sets can have on the performance of the solvers. We do not
declare a solver as the winner based on any ranking by the numbers of solved
instances.

Table~\ref{strat:sampling:solved} shows the
rankings of the solvers for each of the seven randomly sampled benchmark sets
according to the numbers of solved instances.
Rather surprisingly, the rankings
are identical for all seven benchmark sets. This also applies to the rankings
by the penalized average runtime (PAR10) as shown in
Table~\ref{strat:sampling:times}, where runs that timed out after 200 seconds
were penalized with $10 \cdot 200$ seconds.

Two factors that might have contributed to the rankings in
Tables~\ref{strat:sampling:solved} and~\ref{strat:sampling:times} are the
relatively small time out of 200 seconds, and the stratified sampling that
makes all the sampled benchmark sets fairly similar to each other. With longer
time outs we expect to see more variation. The stratified sampling avoids that
instances of certain benchmark families are overrepresented in the final
set. Furthermore, solvers that perform particularly well on certain families
no longer have an advantage when running on benchmark sets where the selection
of instances is biased towards that family. Hence multiple solver runs on
benchmark sets that were sampled in a stratified way together with 
different runtime cutoffs might help to obtain an
unbiased ranking of solvers and finally to declare a winner in a competitive
setting.

%%%%%%%%%%%%%%%%%%%%%%%%%%%%%%%%%%%%%%%%%%%%%%%%%%%%%%%%%%%%%%%%%%%%%%%%%%%%%%%%
%%%%%%%%%%%%%%%%%%%%%%%%%%%%%%%%%%%%%%%%%%%%%%%%%%%%%%%%%%%%%%%%%%%%%%%%%%%%%%%%

\subsection{Showcase: Preprocessing}
\label{sec:prepro}

The purpose of this showcase was to find out how many instances can be
solved solely by preprocessing and to analyze the effects of preprocessing
on the performance of solvers. The latter is closely related to the showcase
on solving (Section~\ref{sec:solving}). All experiments in this showcase were
run on a 64-bit Linux Ubuntu 12.04 
system with four 2.6 GHz 12-core AMD Opteron 6238 CPUs and 
$512$ GB memory in total. The concrete memory limits 
varied for the different experiments.

We carried out experiments in two settings: in the first setting, we ran each
of the four submitted preprocessors shown in Table~\ref{tab:tools}
individually on a given set of original instances. Then, we compared the sets
of instances that were solved by a particular preprocessor. In this
experiment, the preprocessors do not interfere with each other, which allows
to analyze their individual strengths.

In the second setting, we ran the four preprocessors incrementally in multiple
rounds. For example, first preprocessor A is run, and its output, i.e.~the
preprocessed formula, is forwarded to preprocessor B, the output of B in turn
is forwarded to C. Finally D is run on the output of C. Then a new round
starts with A,B,C, and D. In this experiment the individual
preprocessors influence each other, and hence their combined strengths can be
analyzed. Given the number of available preprocessors, there multiple
execution sequences like ABCD, ABDC, AABBCCDD, etc. We aimed at a comprehensive
evaluation by considering as many execution sequences as possible given the
available computational resources. When choosing the execution sequences, we
also took the characteristics of the preprocessors into account. For example,
it might be beneficial to run \preproA, which performs unit literal detection,
before preprocessors that modify the structure of the formula as structure
modifications might be prohibitive for the detection of unit literals.

The time limits used in this showcase were smaller than the ones used in the
other showcases. The choice of the time limits for preprocessing was based on
the conjecture that if an instance can be solved solely by preprocessing, then
it can be solved rather quickly.

\subsubsection{Individual Preprocessing} \label{sec:prepro:indiv}

First, we address the question of how many instances can be solved solely by the
individual preprocessors. Table~\ref{prepro_individual} shows the results of
running the four preprocessors on several benchmark sets described in
Section~\ref{sec:benchmarks}. In these experiments, we used a wall-clock time
limit of 300 seconds and a memory limit of $7$ GB.
\begin{table}[t!]
\begin{center}
\small
\begin{tabular}{|l|rrr|rrr|rrr|rrr|}
\cline{2-13}
\multicolumn{1}{c}{} & \multicolumn{3}{|c|}{\preproA} & \multicolumn{3}{c|}{\preproB} & \multicolumn{3}{c|}{\preproC} & \multicolumn{3}{c|}{\preproD}  \\
\cline{2-13}
\multicolumn{1}{c}{} & \multicolumn{1}{|r}{t} & s & u & t & s & u & t & s & u & t & s & u  \\
\hline
\texttt{eval2012r2 (345)} & 19 & 0 & 19 & 69 & 33 & 36 & 77 & 35 & 42 & 11 & 3 & 8  \\
\texttt{qbf-hardness (198)} & 0 & 0 & 0 & 49 & 12 & 37 & 51 & 12 & 39 & 12 & 0 & 12 \\
\texttt{sauer-reimer (924)} & 81 & 0 & 81 & 137 & 24 & 113 & 153 & 29 & 124 & 78 & 9 & 69 \\
\texttt{planning-CTE (150)} & 0 & 0 & 0 & 3 & 2 & 1 & 7 & 6 & 1 & 0 & 0 & 0 \\
\texttt{conf.-planning (1750) } & 646 & 0 & 646 & 489 & 11 & 478 & 486 & 12 & 474 & 48 & 0 & 48 \\
\texttt{red.-finding (4608)} & 176 & 0 & 176 & 1496 & 837 & 659 & 1650 & 924 & 726 & 674 & 326 & 348 \\
\hline
\end{tabular}
\end{center}
\caption{Total numbers of instances solved by the four considered
  preprocessors (columns~$t$), and solved satisfiable (columns $s$) and
  unsatisfiable instances (columns $u$). Each preprocessor was run
  individually on the benchmark sets. Hence the preprocessors did not influence each other.}
\label{prepro_individual}
\end{table}

Given the statistics in Table~\ref{prepro_individual}, the performance of the
preprocessors varies with respect to the benchmark set. For example, in the
benchmark set \texttt{conformant-planning}, \preproA solves the largest number
of instances whereas \preproC solves the largest number of instances in the
set \texttt{reduction-finding}. Note that by construction \preproA, unlike the other
preprocessors, can only solve unsatisfiable formulas, since it
does not apply variable elimination.

Table~\ref{prepro_indiv_total_sum} shows a combination of the statistics from
Table~\ref{prepro_individual}: an instance is considered to be solved if it
was solved by at least one of the four considered
preprocessors. Interestingly, the total counts in
Table~\ref{prepro_indiv_total_sum} are not always clearly higher than the
largest individual count from Table~\ref{prepro_individual}. This indicates
that there are preprocessors that, regarding their effects, subsume other
preprocessors on certain benchmark sets. For example, in the set
\texttt{planning-CTE}, all instances that are solved by \preproB are also
solved by \preproC.
\begin{table}[ht]
\begin{center}
\small
\begin{tabular}{|l|rrr|}
\cline{2-4}
\multicolumn{1}{c}{} & \multicolumn{3}{|c|}{$A+B+C+D$}  \\
\cline{2-4}
\multicolumn{1}{c}{} & \multicolumn{1}{|r}{t} & s & u \\
\hline
\texttt{eval2012r2} & 87 & 36 & 51 \\
\texttt{qbf-hardness} & 51 & 12 & 39 \\
\texttt{sauer-reimer} & 158 & 29 & 129 \\
\texttt{planning-CTE} & 7 & 6 & 1 \\
\texttt{conf.-planning} & 757 & 12 & 745 \\
\texttt{red.-finding} & 1679 & 940 & 739 \\
\hline
\end{tabular}
\end{center}
\caption{Total numbers of instances solved by any individual preprocessor from
Table~\ref{prepro_individual} (column~$t$), total solved satisfiable
(column $s$) and solved unsatisfiable instances (column $u$). In the
column header, ``A'' labels \preproA, ``B'' labels \preproB, ``C'' labels
\preproC, and ``D'' labels \preproD.}
\label{prepro_indiv_total_sum}
\end{table}
Further results including pairwise comparisons of the individual preprocessors
can be found at the website of the QBF Gallery
2013.\footnote{\url{http://www.kr.tuwien.ac.at/events/qbfgallery2013/results-solving.html}}

\subsubsection{Incremental Preprocessing}

Motivated by the diverse performance of the individual preprocessors
illustrated in the previous section, we investigate whether their
individual strengths can be combined by incremental preprocessing. 
To this end, the preprocessors are run in multiple rounds. In our setting, at
most six rounds were run for each instance. In each round, the instance preprocessed by one
preprocessor is forwarded to another and hence the preprocessors influence
each other.
If an instance is solved by either preprocessor then the whole run
terminates. In the following, ``A'' labels \preproA, ``B'' labels \preproB, ``C'' labels
\preproC, and ``D'' labels \preproD.

A time limit of 120 seconds was imposed for each individual run of A, B, C,
and D. Hence in total, given four preprocessors and six rounds, for each
instance we allowed at most 2880 seconds for preprocessing. This time out is
much larger than the time out of 900 seconds we chose in the showcases on
solving and applications. Our motivation for the showcase
on preprocessing was to analyze the power of preprocessing decoupled from
solving.  Therefore, we decided to allow more time for preprocessing than in a
typical setting where a solver is combined with a preprocessor.

If a preprocessor fails to process
an instance within the given time limit or if it fails due to any
other reason, then its input formula is passed on to the next
preprocessor in the execution sequence without any modifications. 
We considered the benchmark set \texttt{eval2012r2} and tested all $4! = 24$
possible execution sequences of A, B, C, and D.
\begin{table}[t!]
\begin{center}
\small
\begin{tabular}{|l|rrr|}
\cline{2-4}
\multicolumn{1}{c}{} & \multicolumn{3}{|c|}{\texttt{eval2012r2}}  \\
\cline{2-4}
\multicolumn{1}{c}{} & \multicolumn{1}{|c}{$t$} & $s$ & $u$ \\
\hline
$(\mathit{ABCD})^{6}$ & 107 & 44 & 63 \\
$(\mathit{ABDC})^{6}$ & 106 & 42 & 64 \\
$(\mathit{ACBD})^{6}$ & 103 & 43 & 60 \\
$(\mathit{ACDB})^{6}$ & 103 & 43 & 60 \\
$(\mathit{ADBC})^{6}$ & 103 & 41 & 62 \\
$(\mathit{ADCB})^{6}$ & 102 & 41 & 61 \\

$(\mathit{BACD})^{6}$ & 102 & 41 & 61 \\
$(\mathit{BADC})^{6}$ & 102 & 41 & 61 \\
$(\mathit{BCAD})^{6}$ & 101 & 41 & 60 \\
$(\mathit{BCDA})^{6}$ & 103 & 42 & 61 \\
$(\mathit{BDAC})^{6}$ & 101 & 41 & 60 \\
$(\mathit{BDCA})^{6}$ & 99 & 39 & 60 \\

$(\mathit{CABD})^{6}$ & 99 & 41 & 58 \\
$(\mathit{CADB})^{6}$ & 99 & 40 & 59 \\
$(\mathit{CBAD})^{6}$ & 98 & 40 & 58 \\
$(\mathit{CBDA})^{6}$ & 98 & 40 & 58 \\
$(\mathit{CDAB})^{6}$ & 100 & 41 & 59 \\
$(\mathit{CDBA})^{6}$ & 100 & 40 & 60 \\

$(\mathit{DABC})^{6}$ & 102 & 38 & 64 \\
$(\mathit{DACB})^{6}$ & 100 & 38 & 62 \\
$(\mathit{DBAC})^{6}$ & 101 & 38 & 63 \\
$(\mathit{DBCA})^{6}$ & 100 & 38 & 62 \\
$(\mathit{DCAB})^{6}$ & 96 & 37 & 59 \\
$(\mathit{DCBA})^{6}$ & 96 & 37 & 59 \\
\hline
\multicolumn{1}{|c|}{VBS} & 119  & 48 & 71 \\
\hline
\end{tabular}
\end{center}
\caption{ Numbers of instances solved by one out of 24 possible execution
  sequences of A, B, C, and D within at most six rounds (column~$t$), solved satisfiable (column $s$),
and unsatisfiable instances (column $u$), where ``A'' labels
\preproA, ``B'' labels \preproB, ``C'' labels \preproC, and ``D'' labels
\preproD. The results of the virtual best solver (VBS) is shown in the last 
line.}
\label{prepro_inc_eval2012r2_all_sequences}
\end{table}

Table~\ref{prepro_inc_eval2012r2_all_sequences} shows the number of
instances solved by each execution sequence. Each execution sequence
solves more instances than any of the individual preprocessors
(Tables~\ref{prepro_individual} and~\ref{prepro_indiv_total_sum}). In
total, 119 instances where solved by any of the execution sequences,
which is 34\% of the instances contained in the benchmark set
\texttt{eval2012r2}. With individual preprocessing, in total 87
instances (25\%) were solved by any of the preprocessors (first line
in Table~\ref{prepro_indiv_total_sum}). These statistics clearly
indicate the benefits of incremental preprocessing in terms of solved
instances.

However, the performance of incremental preprocessing is sensitive to
the ordering of the preprocessors in an execution sequence. For
example, the sequences ABCD and ABDC with the prefix AB solve the
largest number of instances (107 and 106, respectively). In contrast
to that, the sequences DCAB and DCBA with the prefix DC solve the
smallest number of instances (96 each). This difference indicates that
the techniques implemented in individual preprocessors might have
a negative effect in incremental preprocessing. One preprocessor might
destroy the structure of the formula, which in turn might restrict the
effects of another preprocessor relying on that structure.

According to the results shown in
Table~\ref{prepro_inc_eval2012r2_all_sequences}, the largest number of
instances in the \texttt{eval2012r2} were solved with the sequence
ABCD. Based on this observation, we ran the sequence ABCD on the other
benchmark sets for at most six rounds.  Table~\ref{prepro_inc} shows
the results of these experiments. Except for the set
\texttt{qbf-hardness}, incremental preprocessing solves considerably
more instances than the individual preprocessors
(Table~\ref{prepro_indiv_total_sum}). For example, for the sets
\texttt{planning-CTE} and \texttt{conformant-planning}, 57\% and 23\% more
instances are solved, respectively.
\begin{table}[ht]
\begin{center}
\small
\begin{tabular}{|l|rrr|}
\cline{2-4}
\multicolumn{1}{c}{} & \multicolumn{3}{|c|}{$(\mathit{ABCD})^{6}$}  \\
\cline{2-4}
\multicolumn{1}{c}{} & \multicolumn{1}{|r}{t} & s & u \\
\hline
\texttt{eval2012r2} & 107 & 44 & 63 \\
\texttt{qbf-hardness} & 51 & 12 & 39 \\
\texttt{sauer-reimer} & 180 & 31 & 149 \\
\texttt{planning-CTE} & 11 & 8 & 3 \\
\texttt{conf.-planning} & 938 & 13 & 925 \\
\texttt{red.-finding} & 1855 & 936 & 919 \\
\hline
\end{tabular}
\end{center}
\caption{Incremental preprocessing by running the execution sequence ABCD for at most six rounds. Total
numbers of solved instances (columns~$t$), solved satisfiable (columns $s$),
and unsatisfiable instances (columns $u$), where ``A'' labels
\preproA, ``B'' labels \preproB, ``C'' labels \preproC, and ``D'' labels
\preproD.}
\label{prepro_inc}
\end{table}

In an additional experiment, we tested selected execution sequences
from Table~\ref{prepro_inc_eval2012r2_all_sequences} on the benchmark
set \texttt{eval2012r2}, where each preprocessor is run twice in a
row. We selected the sequences to be tested according to the numbers
of solved instances shown in
Table~\ref{prepro_inc_eval2012r2_all_sequences} and the individual
characteristics of the preprocessors.

As in the previous experiments, we used a wall-clock time limit of 120
seconds for each individual call of a preprocessor and at most six
rounds of incremental preprocessing. The results in
Table~\ref{prepro_inc_subsequent_calls} show a moderate increase in
the number of solved instances, except for the sequences
$(B^2C^2D^2A^2)^{6}$ and $(D^2A^2B^2C^2)^{6}$. These observations
indicate that after some time preprocessing reaches a point where
little or no progress at all is made. In the following, we
analyze situations of this kind.

\begin{table}[ht]
\begin{center}
\small
\begin{tabular}{|l|rrr|}
\cline{2-4}
\multicolumn{1}{c}{} & \multicolumn{3}{|c|}{\texttt{eval2012r2}}  \\
\cline{2-4}
\multicolumn{1}{c}{} & \multicolumn{1}{|r}{$t$} & $s$ & $u$ \\
\hline
$(A^2B^2C^2D^2)^{6}$ & 111 & 46 & 65 \\
$(A^2B^2D^2C^2)^{6}$ & 111 & 45 & 66 \\
$(A^2D^2B^2C^2)^{6}$ & 104 & 42 & 62 \\
$(B^2C^2D^2A^2)^{6}$ & 103 & 42 & 61 \\
$(D^2A^2B^2C^2)^{6}$ & 102 & 38 & 64 \\
\multicolumn{1}{|c|}{Total} & 115  & 48 & 67 \\
\hline
\end{tabular}
\end{center}
\caption{Incremental preprocessing with different execution
  sequences where preprocessors are called twice. For example, the string
  $(A^2B^2C^2D^2)^{6}$ indicates the execution sequence AABBCCDD where at most
  six rounds are run. The table shows the total
numbers of solved instances (columns~$t$), solved satisfiable (columns $s$),
and unsatisfiable instances (columns $u$), where ``A'' labels
\preproA, ``B'' labels \preproB, ``C'' labels \preproC, and ``D'' labels
\preproD.}
\label{prepro_inc_subsequent_calls}
\end{table}

\subsubsection{Detection of Fixpoints}

When running incremental preprocessing in multiple rounds, it might
happen that an instance is not modified anymore during a round. In
this case, a fixpoint has been reached and hence preprocessing can be
stopped.

In the experimental evaluation of incremental preprocessing, we
implemented the detection of fixpoints as follows. At the beginning of
each round, before the first preprocessor in the execution sequence is
run, the clause set of the current instance is
normalized. Normalization discards tautological clauses and sorts the
literals of each clause in the set. Then the set of clauses is sorted
using the Linux command line tool \texttt{sort}. An MD5 hash value is
computed for this normalized instance using the Linux command line
tool \texttt{openssl}. The normalized instance is used only for the
computation of the hash value and it is not forwarded to the
preprocessors. Hence the detection of fixpoints does not interfere
with preprocessing. At the end of the current round, after the last
preprocessor in the execution sequence has been run, a hash value is
computed for the normalized clause set of the instance produced by the
last preprocessor. If the hash values at the beginning and at the end
of a round are equal then the clause set was not modified by
preprocessing in the current round. Hence a fixpoint has been reached
and the run terminates.

Due to normalization as described above, the detection of fixpoints we
implemented does not distinguish between instances that differ in terms of
the ordering of clauses or the ordering of the literals in the clauses.  However, in practice
different orderings might have an impact on the performance of the
preprocessors as the heuristics internal to a preprocessor might be
influenced. In the experimental analysis, we did not analyze the
effects of different orderings of clauses or literals.

Table~\ref{prepro_inc_fixpoints_and_solved_combined} shows statistics
on fixpoints and solved instances in each out of six rounds when
running the five execution sequences from
Table~\ref{prepro_inc_subsequent_calls} on the set
\texttt{eval2012r2}. The vast majority of instances is solved already
in the first round. No instances are solved in rounds five and
six. The number of fixpoints decreases considerably from round four
up to round six. This indicates that it is justified to run a limited
number of rounds of incremental preprocessing. For example, in a
related experiment (not shown in the tables) where we ran the
execution sequence $\mathit{ABCD}$ in at most 12 rounds on the set
\texttt{eval2012r2}, no instance was solved in rounds 7--12.
Likewise, when increasing the number of rounds to 24, then no
instance was solved in rounds 7--24.
\begin{table}[t!]
\begin{center}
\small
\begin{tabular}{|l|rrrrrrr|rrrrrrr|}
\cline{2-15}
\multicolumn{1}{c}{} & \multicolumn{14}{|c|}{\texttt{eval2012r2}}  \\
\cline{2-15}
\multicolumn{1}{c}{} & \multicolumn{7}{|c}{\emph{Solved Instances}} &
\multicolumn{7}{|c|}{\emph{Detected Fixpoints}} \\
\multicolumn{1}{c}{} & \multicolumn{1}{|c}{1} & 2 & 3 & 4 & 5 & 6 & \multicolumn{1}{c}{$\Sigma$} &
\multicolumn{1}{|c}{1} & 2 & 3 & 4 & 5 & 6 & \multicolumn{1}{c|}{$\Sigma$} \\
\hline
$(A^2B^2C^2D^2)^{6}$ & 100 & 8 & 2 & 1 & 0 & 0 & 111 & 2 & 52 & 76 & 60 & 14 & 2 & 206 \\
$(A^2B^2D^2C^2)^{6}$ & 101 & 7 & 2 & 1 & 0 & 0 & 111 & 1 & 16 & 113 & 58 & 16 & 3 & 207 \\
$(A^2D^2B^2C^2)^{6}$ & 96 & 8 & 0 & 0 & 0 & 0 & 104 & 0 & 16 & 137 & 53 & 12 & 0 & 218 \\
$(B^2C^2D^2A^2)^{6}$ & 96 & 5 & 1 & 1 & 0 & 0 & 103 & 2 & 46 & 81 & 66 & 14 & 2 & 211 \\
$(D^2A^2B^2C^2)^{6}$ & 93 & 8 & 1 & 0 & 0 & 0 & 102 & 2 & 12 & 141 & 49 & 10 & 4 & 218 \\
\hline
\end{tabular}
\end{center}
\caption{Numbers of instances solved and fixpoints detected in each of six rounds (columns ``1,\ldots,6'') and the total
  number of solved instances and fixpoints (columns '$\Sigma$') for the five execution
sequences from Table~\ref{prepro_inc_subsequent_calls}.}
\label{prepro_inc_fixpoints_and_solved_combined}
\end{table}

\subsubsection{Solving Performance of Preprocessors}
\label{sec:prepro:solving}

In contrast to the experiments conducted in the case of the 
best-foot-forward experiments above, in the following we are interested
in the effects of applying different combinations of preprocessors
in multiple rounds and assess the solving performance of preprocessors.
As illustrated by 
Table~\ref{prepro_inc_fixpoints_and_solved_combined}, the numbers of
instances solved by preprocessing using a particular execution
sequence is sensitive to the ordering of the preprocessing tools in
the sequence. To further analyze this effect, we tested the
combination of incremental preprocessing and solving. Thereby, we
preprocessed the benchmark set \texttt{eval2012r2} (345 instances)
using the execution sequences $(A^2B^2C^2D^2)^{6}$ and
$(A^2D^2B^2C^2)^{6}$. This way, we obtained the two new benchmark sets
\texttt{AABBCCDD} (234 instances
remaining unsolved after preprocessing) and
\texttt{AADDBBCC} (241 instances
remaining) listed in Section~\ref{sec:benchmarks},
respectively. We selected the sequence $(A^2B^2C^2D^2)^{6}$ because it solved
the largest number of instances
(Table~\ref{prepro_inc_fixpoints_and_solved_combined}) and 
because $(ABCD)^{6}$ performed best according to Table~\ref{prepro_inc_eval2012r2_all_sequences}. Since \bloqqer (``B'')
and \squeezebf (``D'') have different characteristics,  we additionally 
selected the sequence $(A^2D^2B^2C^2)^{6}$ where the ordering of these two
preprocessors is swapped and still \preproB is executed before
\preproC (``C''). We did not consider the sequence $(A^2D^2C^2B^2)^{6}$ where
only \preproB and \preproD are swapped since, according to the results
shown in Table~\ref{prepro_inc_eval2012r2_all_sequences}, the execution 
ordering $(ADCB)$ solved one instance less than the execution ordering 
$(ADBC)$ in the sequence $(A^2D^2B^2C^2)^{6}$ we selected.

Tables~\ref{tab_AABBCCDDeval12_solving}
and~\ref{tab_AADDBBCCeval12_solving} show the performance of solvers
on the two benchmark sets. 
The different rankings of the solvers in the tables indicate that their 
performance is sensitive to the execution ordering of the 
preprocessors. Furthermore, the total number of instances solved
by preprocessing \emph{and} by solving is different for the two benchmark
sets. For the set \texttt{AABBCCDD},
111 instances were solved by preprocessing (first line in
Table~\ref{prepro_inc_fixpoints_and_solved_combined}) and 92 instances
were solved by the best solver (first line in
Table~\ref{tab_AABBCCDDeval12_solving}), giving a total of 203 solved
instances. For the set
\texttt{AADDBBCC}, 104 instances were
solved by preprocessing (third line in
Table~\ref{prepro_inc_fixpoints_and_solved_combined}) and 104 by the
best solver (first line in Table~\ref{tab_AADDBBCCeval12_solving}),
which gives 208 solved instances in total. That is, although
preprocessing alone using the execution sequence $(A^2D^2B^2C^2)^{6}$
solves fewer instances than when using the sequence
$(A^2B^2C^2D^2)^{6}$, solving performs better on the 
instances that were preprocessed using the former and results in a higher total number of instances solved by preprocessing \emph{and} solving.
\begin{table}[ht] \small
\centering
 \begin{tabular}{|l|r|r|r|r|r|r|}
\hline
&\multicolumn{4}{c|}{\textbf{number of solved formulas}} & \multicolumn{2}{c|}{\textbf{runtime (sec)}}\\\hline
\textbf{solver}&solved&sat&unsat&unique&avg&total\\
\hline
\rareqs&92&57&35&8&35&131K\\
\ldepqbf&90&52&38&0&87&137K\\
\depqbfm&87&50&37&0&98&140K\\
\ooqm&79&49&30&4&107&145K\\
\hiqqer&78&45&33&0&79&147K\\
\dooqm&74&48&26&0&93&147K\\
\bghostc&64&46&18&0&127&160K\\
\ghostc&64&46&18&0&124&149K\\
\ghost&57&43&14&0&83&164K\\
\sdooq&54&37&17&0&103&167K\\
\nenofexm&48&32&16&3&59&170K\\
\sqube&46&33&13&1&88&173K\\
\hline
 \end{tabular}
 \caption{Related to
   Figure~\ref{fig_prepro_inc_subsequent_calls_solving_cactus_plot}:
   solver performance on the set
   \texttt{AABBCCDD} (234 instances)}
 \label{tab_AABBCCDDeval12_solving}
\end{table}
\begin{table}[ht] \small
\centering
 \begin{tabular}{|l|r|r|r|r|r|r|}
\hline
&\multicolumn{4}{c|}{\textbf{number of solved formulas}} & \multicolumn{2}{c|}{\textbf{runtime (sec)}}\\\hline
\textbf{solver}&solved&sat&unsat&unique&avg&total\\
\hline
\ldepqbf&104&59&45&0&86&132K\\
\rareqs&104&59&45&7&54&128K\\
\depqbfm&102&59&43&0&87&134K\\
\hiqqer&90&52&38&0&99&144K\\
\ooqm&90&56&34&2&61&141K\\
\dooqm&80&54&26&0&81&151K\\
\sdooq&64&43&21&0&101&165K\\
\ghost&61&43&18&0&92&168K\\
\bghostc&59&40&19&0&77&168K\\
\ghostc&59&40&19&0&78&168K\\
\sqube&58&37&21&0&56&168K\\
\nenofexm&52&34&18&3&53&172K\\
\hline
 \end{tabular}
 \caption{Related to
   Figure~\ref{fig_prepro_inc_subsequent_calls_solving_cactus_plot}:
   solver performance on the set
   \texttt{AADDBBCC} (241 instances)}
 \label{tab_AADDBBCCeval12_solving}
\end{table}

\begin{figure}[t]
\centering
\includegraphics[scale=0.54]{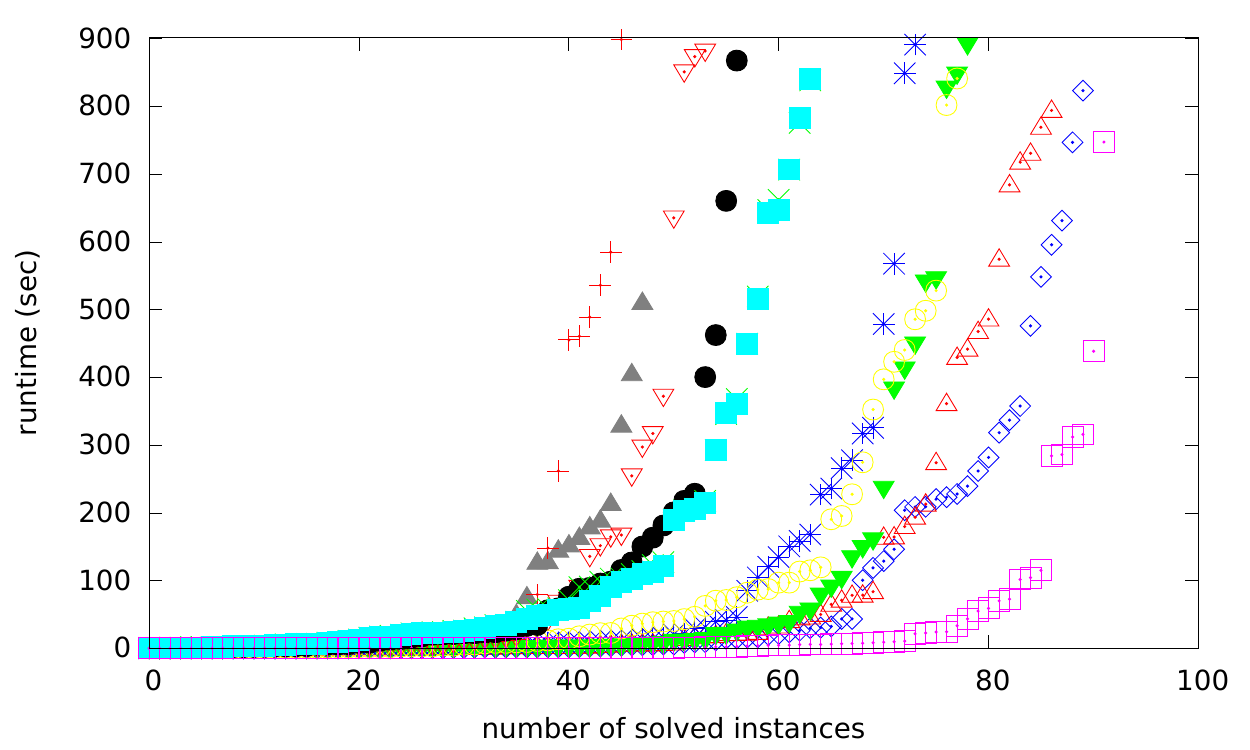}
\includegraphics[scale=0.54]{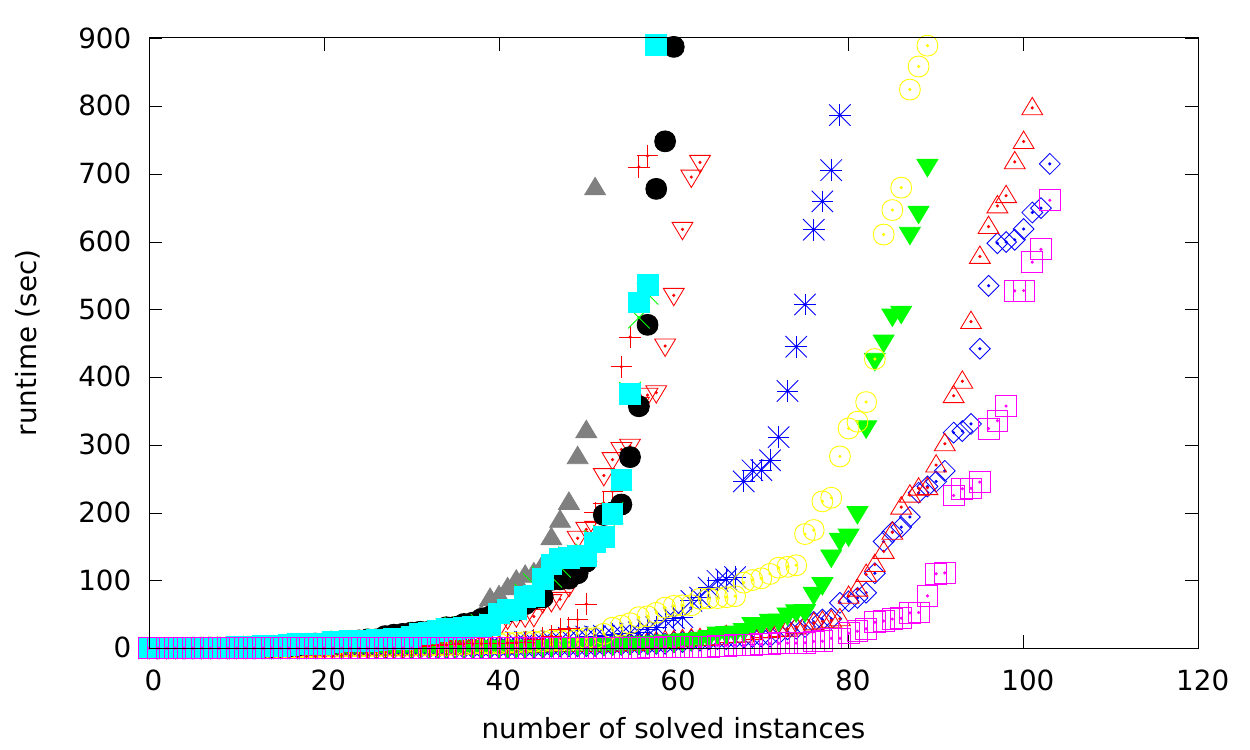}
\includegraphics{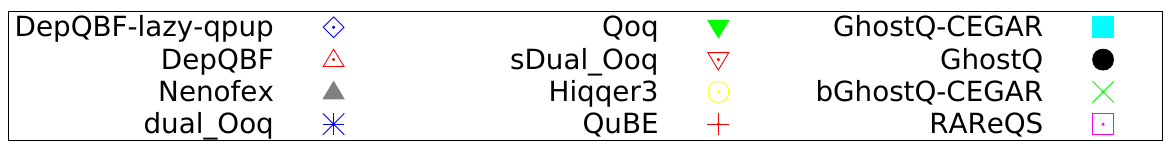}

\caption{Related to Tables~\ref{tab_AABBCCDDeval12_solving}
  and~\ref{tab_AADDBBCCeval12_solving}: solver performance on the sets
  \texttt{AABBCCDD} (left) and
  \texttt{AADDBBCC} (right). These sets
  were obtained from the set \texttt{eval2012r2} by preprocessing
  using the execution sequences $(A^2B^2C^2D^2)^{6}$ and
  $(A^2D^2B^2C^2)^{6}$, respectively.}
\label{fig_prepro_inc_subsequent_calls_solving_cactus_plot}
\end{figure}

%%%%%%%%%%%%%%%%%%%%%%%%%%%%%%%%%%%%%%%%%%%%%%%%%%%%%%%%%%%%%%%%%%%%%%%%%%%%%%%%
%%%%%%%%%%%%%%%%%%%%%%%%%%%%%%%%%%%%%%%%%%%%%%%%%%%%%%%%%%%%%%%%%%%%%%%%%%%%%%%%

\subsection{Showcase: Applications}
\label{sec:apps}

The purpose of the showcase on QBF applications was to evaluate the benchmark
families submitted by the participants. The goal was to find out which types of
solvers perform well on a specific family, what the reasons are for good or
bad performance, and to identify future research directions to improve QBF
solvers for benchmark families that arise from practical applications.

From the benchmark sets listed in Section~\ref{sec:benchmarks}, the following
five sets are related to practical applications: \texttt{reduction-finding},
\texttt{conformant-planning}, \texttt{planning-CTE}, \texttt{sauer-reimer},
and \texttt{qbf-hardness}. We split the set \texttt{conformant-planning} into
the two subsets \texttt{conformant-plan\-ning-bomb} and
\texttt{conformant-planning-dungeon}, containing instances with different
characteristics. All the formulas considered in this showcase were newly 
submitted to the QBF Gallery.  They
have not been used in an evaluation before and are not available from QBFLIB.

From the resulting six sets of application-related benchmarks, we randomly
sampled 150 formulas each and tested the submitted solvers on each of these
sampled sets. In the following experiments, a time limit of 900 seconds and a
memory limit of $7$ GB was used. We did not consider preprocessing in order to
evaluate the solvers on the original instances as they were generated by the
participants.

\begin{figure}[t!]

\centering
\begin{tikzpicture}
\node (a) {\includegraphics[scale=0.6]{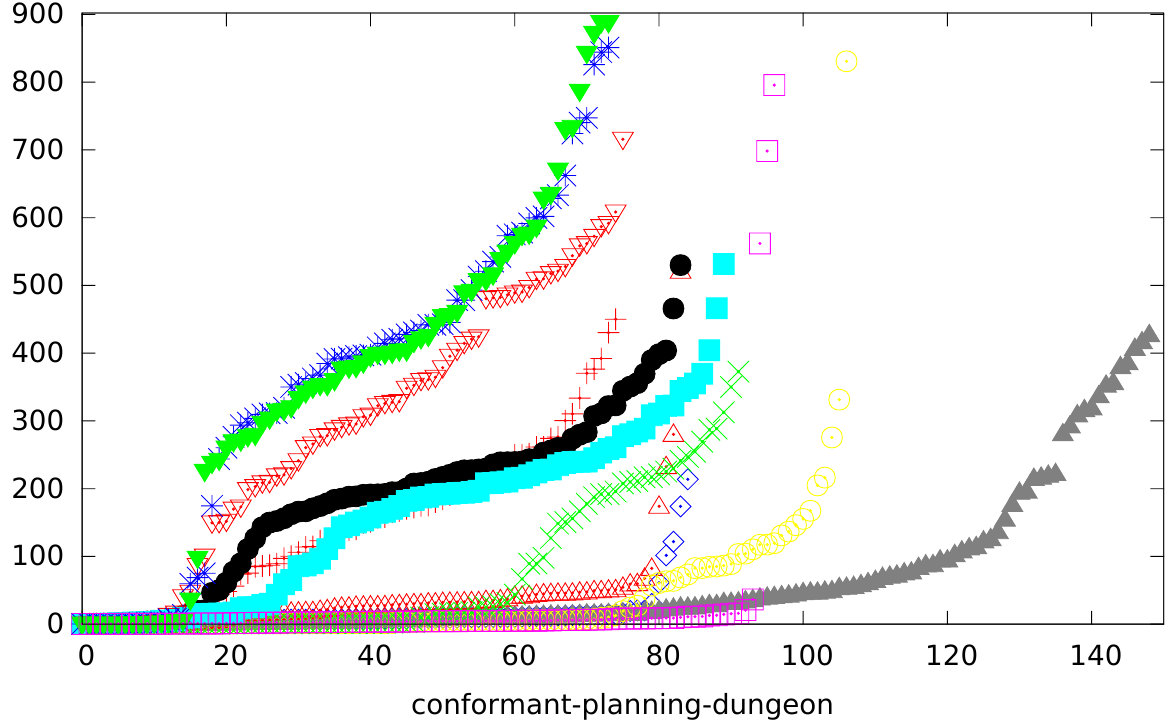}
\includegraphics[scale=0.6]{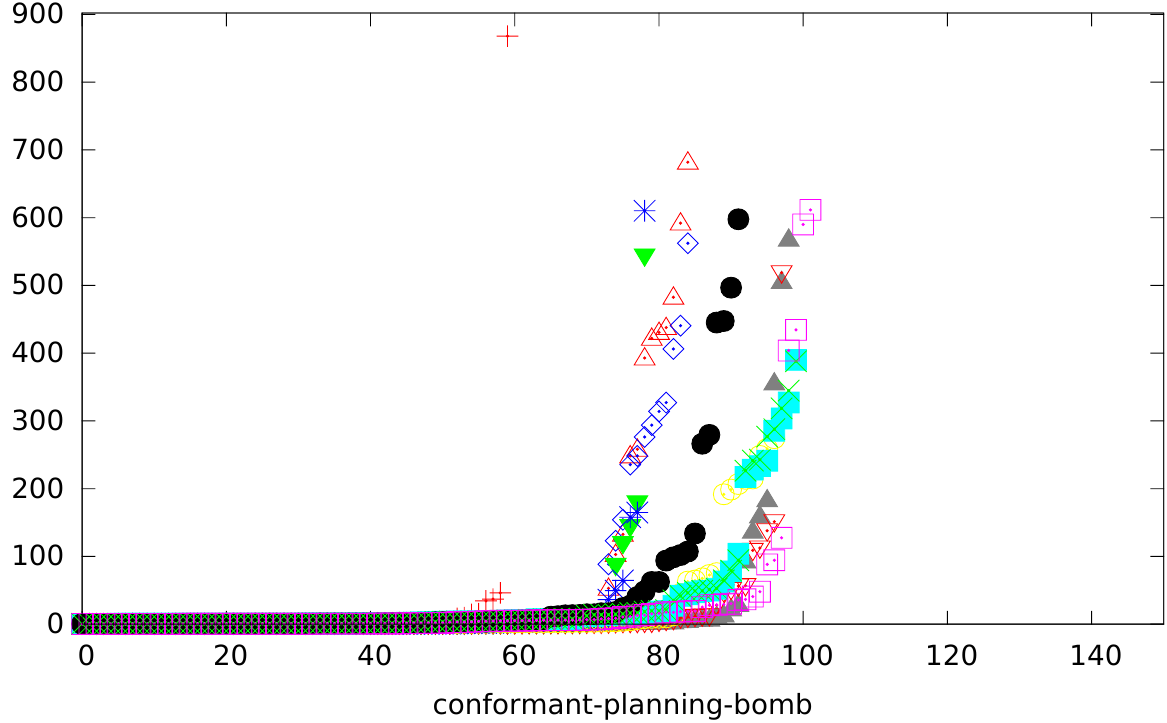}
};
\node (b) [yshift=-3.2cm,below of=a] {
\includegraphics[scale=0.6]{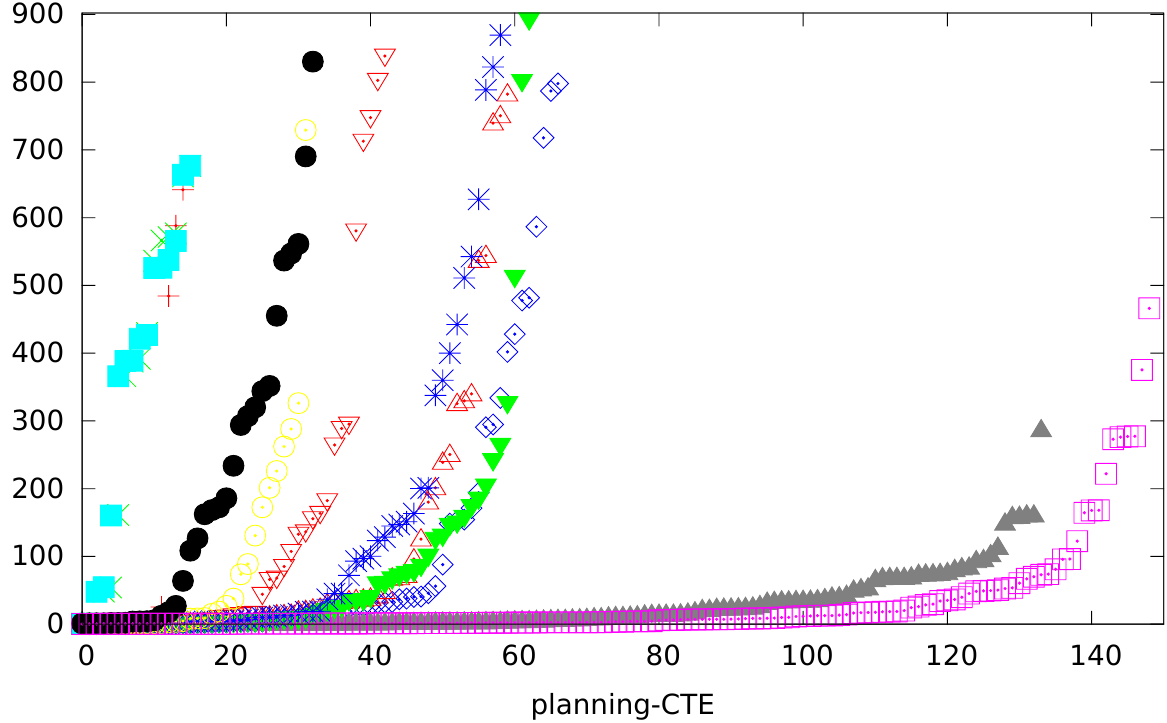}
\includegraphics[scale=0.6]{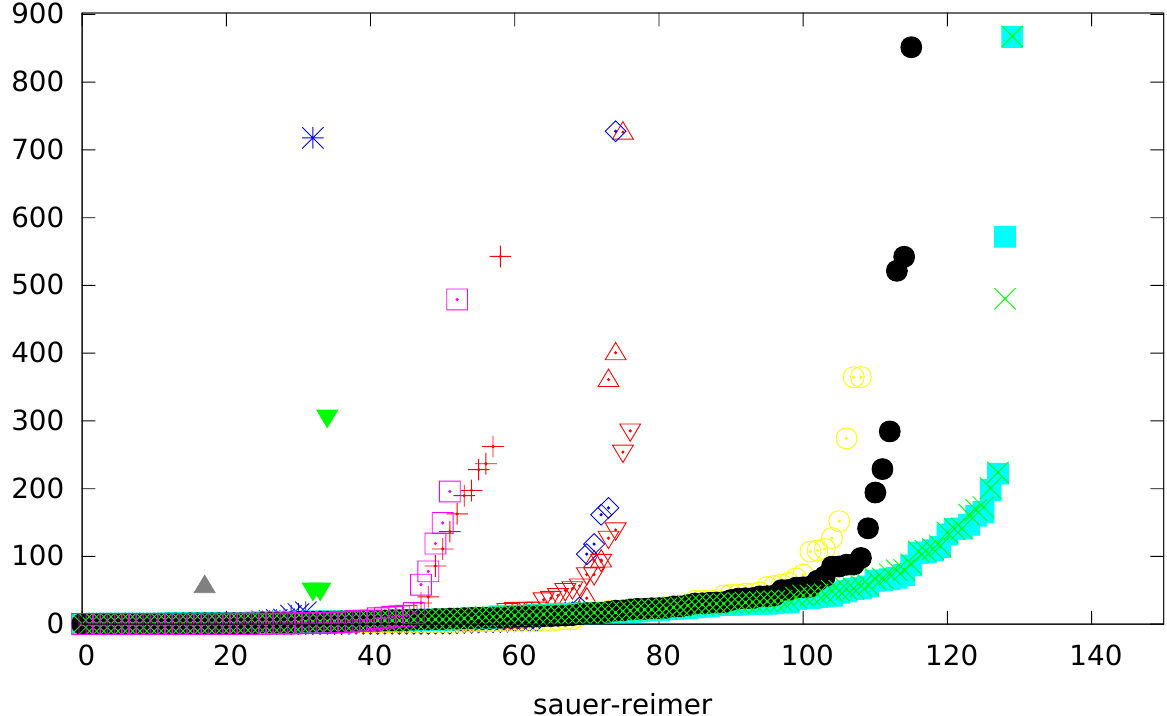}
};
\node (c) [yshift=-3.2cm,below of=b] {
\includegraphics[scale=0.6]{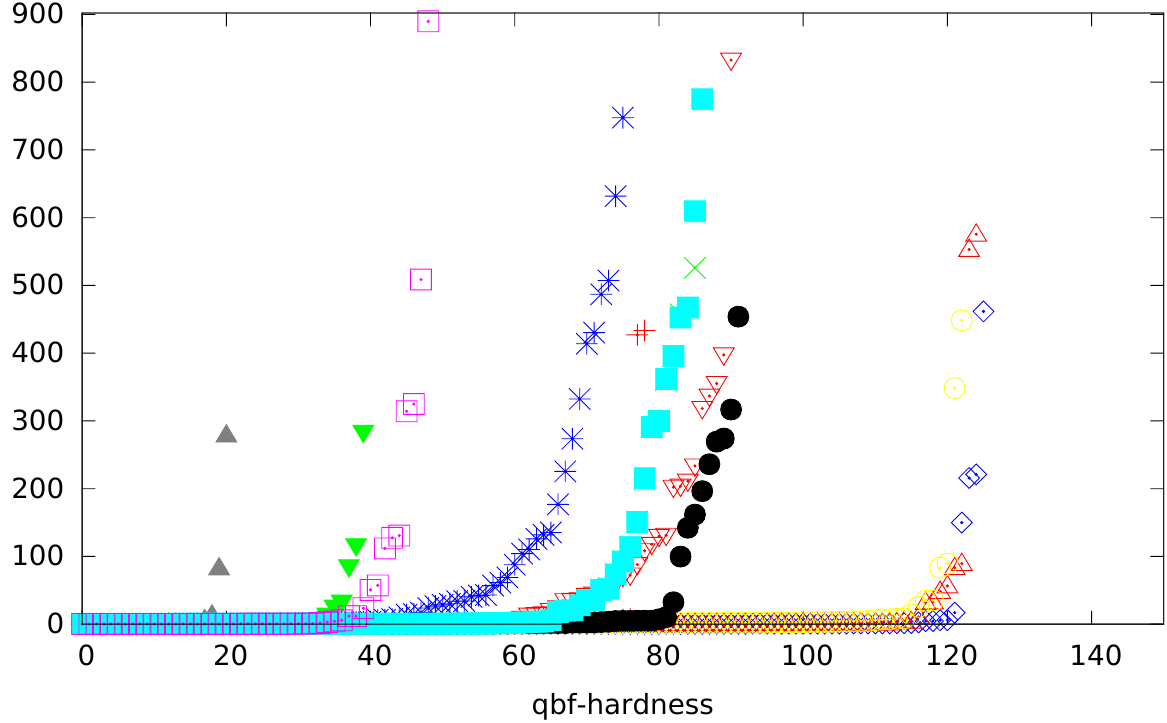}
\includegraphics[scale=0.6]{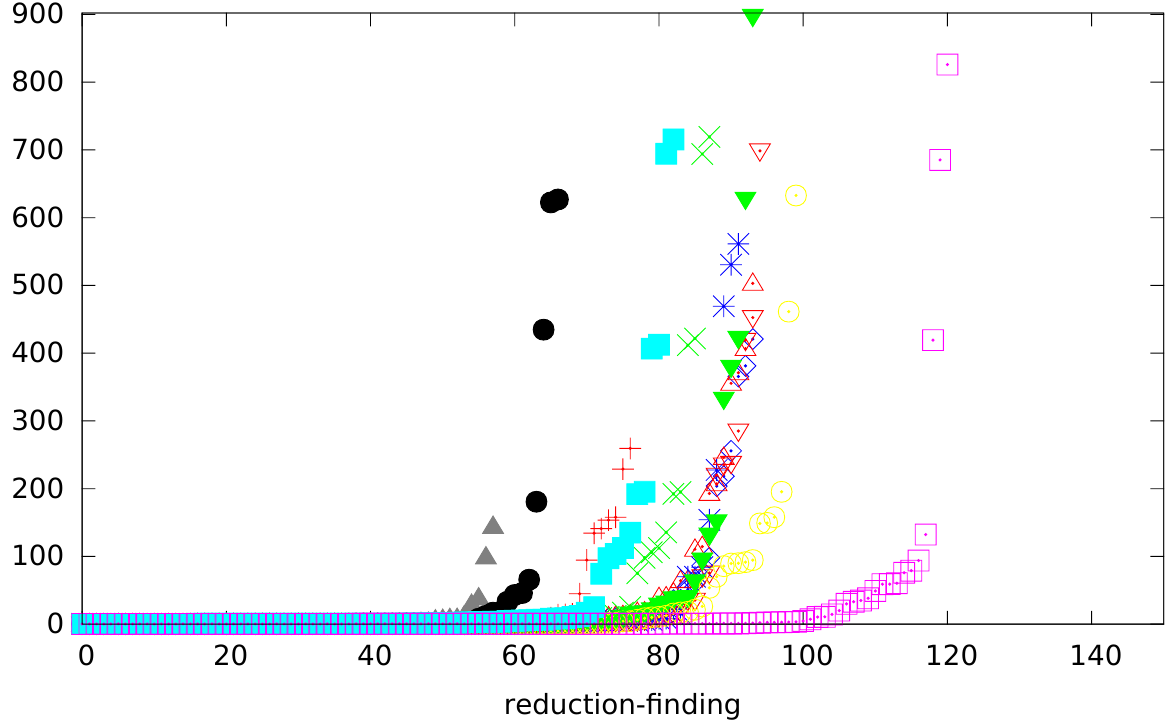}};
\node [below of = b,yshift=-5.5cm]{\footnotesize{\textsf{number of solved instances}}};
\node [yshift=-5.8cm, xshift=-6.5cm,left of = a,rotate=90,xshift=2cm]{\footnotesize{\textsf{runtime (sec)}}};
\end{tikzpicture}
\caption{Runtimes of the solvers on six benchmark sets with 150 formulas each
related to QBF applications (from top left):
\texttt{conformant-planning-dungeon}, \texttt{conformant-planning-bomb},
\texttt{planning-CTE}, \texttt{sauer-reimer}, \texttt{qbf-hardness}, and
\texttt{reduction-finding}. In the plots, each color represents a particular
solver. 
}
\label{fig_applications_cactus_plot}
\end{figure}

Figure~\ref{fig_applications_cactus_plot} shows the run times of the solvers
on each set. The plots indicate that the performance of the solvers greatly
varies with respect to the benchmark set. Tables~\ref{confplanningbomb}
to~\ref{sauerreimer} show detailed solving statistics for each of the
considered benchmark sets, illustrating the different rankings of the solvers
in terms of the numbers of solved formulas. 
 For example, \nenofexm clearly outperforms
the other solvers on the sets \texttt{conformant-planning-dungeon} and
\texttt{planning-CTE} (except \rareqs), but is not competitive on the other
sets. This observation is interesting because \nenofexm and \rareqs rely on
variable expansion. According to the experiments, expansion works particularly
well on the considered formulas related to planning problems. On these
problems, search-based solvers such as \depqbf, \ghostc, and \hiqqer, for
example, perform considerably worse. However,
these solvers perform well on other benchmarks sets like
\texttt{qbf-hardness}.

The diverse solver performance on the different application benchmark families
as illustrated by Figure~\ref{fig_applications_cactus_plot} motivates
exploring potential combinations of the techniques implemented in search-based
solvers and expansion-based solvers in future work. 
The difference in the performance depends
on the considered benchmark family. In that respect, the difference is more pronounced
in the showcase on applications than in the showcase on solving due to the
homogeneity of instances within a particular application benchmark family.

%%%%%%%%%%%%%%%%%%%%%%%%%%%%%%%%%%%%%%%%%%%%%%%%%%%%%%%%%%%%%%%%%%%%%%%%%%%%%%%%
%%%%%%%%%%%%%%%%%%%%%%%%%%%%%%%%%%%%%%%%%%%%%%%%%%%%%%%%%%%%%%%%%%%%%%%%%%%%%%%%

\subsection{Showcase: Certificates}
\label{sec:cert}

The goal of this showcase was to evaluate the current state of the art
of the generation of proofs, certificates, and strategies, which has
been a long standing problem in QBF research.  Proofs, certificates,
and strategies allow to verify the result produced by a QBF solver
independently from the solving process and provide a deeper insight
into the reasons for the (un)satisfiability of a QBF. This insight can
be helpful for QBF applications where a mere ``SAT/UNSAT'' result
produced by the solver is insufficient (e.g., QBF-based synthesis~\cite{DBLP:conf/vmcai/BloemKS14}).

Given an unsatisfiable QBF $\psi$, a
Q-resolution~\cite{DBLP:journals/iandc/BuningKF95} \emph{proof} of
unsatisfiability is a sequence of Q-resolution steps that
demonstrates the derivation of the empty clause from $\psi$. If the
QBF $\psi$ is satisfiable, then a variant of Q-resolution, called \emph{term
resolution}~\cite{DBLP:journals/jair/GiunchigliaNT06,DBLP:conf/tableaux/Letz02,DBLP:conf/cp/ZhangM02},
can be applied to derive the empty cube\footnote{A cube is a
  conjunction of literals.} from $\psi$ by means of a Q-resolution
proof of satisfiability.  Once a proof $\Pi$ (of unsatisfiability or
satisfiability, respectively) has been found for a QBF $\psi$, a
\emph{strategy}~\cite{DBLP:conf/ijcai/GoultiaevaGB11} or a
\emph{certificate}~\cite{DBLP:journals/fmsd/BalabanovJ12} can be
extracted from $\Pi$.

The notion of a \emph{strategy} is related to the game-oriented view
of the semantics of QBF, where the universal and existential player,
who assign the universally and existentially quantified variables, try
to falsify and satisfy the formula, respectively. Thereby, the two
players assign values to the variables in alternating fashion,
starting at the left end of the quantifier prefix. The existential
(universal) player wins if her selection of values satisfies
(falsifies) the formula regardless of the values selected by the other
player. A strategy for a satisfiable (unsatisfiable) QBF represents
the winning choices of values the existential (universal) player must
select depending on the values previously assigned by the universal
(existential) player.

A \emph{certificate} of a satisfiable QBF $\psi$ is a set $F =
\{f_{x_1}(D_{x_1}), \ldots, f_{x_n}(D_{x_n})\}$ of Skolem functions
$f_{x_i}(D_{x_i})$ for the existential variables $x_1, \ldots, x_n$ of
$\psi$. A Skolem function $f_{x_i}(D_{x_i})$ of an existential
variable $x_i$ depends on the universal variables $D_{x_i}$ that
appear to the left of $x_i$ in the quantifier prefix of $\psi$. In the
process of Skolemization, each occurrence of an existential variable
$x_i$ in $\psi$ is replaced by its Skolem function
$f_{x_i}(D_{x_i})$. The QBF $\psi'$ resulting from Skolemization
contains only universal variables and is satisfiable, which can be checked using
a propositional satisfiability (SAT) solver by checking whether the negated formula
$\neg \psi'$ is unsatisfiable. 
Certificates of unsatisfiable QBFs are defined analogously in terms of Herbrand
functions of universal variables. The process of Herbrandization
results in an unsatisfiable formula containing only existential variables.

Compared to strategies, which are based on a game-oriented view, certificates
in terms of Skolem and Herbrand functions 
allow for a more explicit, functional representation of values of 
existential (universal) variables. Apart from that, the concepts of strategies
and certificates are similar.

In this showcase, we used the benchmark set \texttt{eval2012r2}
without preprocessing in order to evaluate the generation of proofs
and certificates on original instances. Due to the lack of submissions
of tools for the generation of strategies, we focused on proofs and
certificates. Since only one proof-producing solver (i.e.~\depqbf) and
only two certificate extraction tools (i.e.~\resqu and \qbfcert) were
officially submitted, we additionally considered further publicly
available tools as shown in the lower part of
Table~\ref{certification_results}.
\begin{table}[t!]
\begin{center}
\begin{tabular}{|ccc|}
\hline
Workflow & Solving+Proof Extr. & Cert.~Extr.+Checking  \\
\hline
\multicolumn{3}{|c|}{\emph{Submitted Tools}} \\
&& \\
\depqbfm~and \qbfcert & 91 (34, 57) & 67 (20, 47) \\
\depqbfm~and \resqu & 91 (34, 57) & 63 (22, 41) \\
\hline
\multicolumn{3}{|c|}{\emph{Additional Tools}} \\
&& \\
\skizzo and \ozziks & 88 (36, 52) & 35 (35, \!\ \ 0) \\
\squolem and \qbv & 38 (19, 19) & 38 (19, 19) \\
\squolem and \resqu & 38 (19, 19) & 19 (\!\ \ 0, 19) \\
\qubecert and \checker & 80 (25, 55) & 32 (11, 21) \\
\qubecert and \resqu & 80 (25, 55) & 52 (17, 35) \\
\hline
\end{tabular}
\end{center}
\caption{Experiments with the generation and checking of proofs and
certificates using various solvers and tools (first column). After an
instances has been solved (second column, numbers of unsatisfiable and
satisfiable instances in parentheses), a certificate is extracted and checked
(third column, numbers of (un)satisfiable instances where a certificate was
successfully extracted and checked). The upper and lower parts of the table,
respectively, show the results obtained with officially submitted tools and
additional, publicly available tools.}
\label{certification_results}
\end{table}

Due to the small number of tools submitted to this showcase, we refrain from
ranking the tools according to their performance. Instead, we comment on the
results of the experiments shown in Table~\ref{certification_results} using
various workflows consisting of different solvers and tools for the
extraction and checking of certificates.

All experiments were run using a wall-clock time limit of 600 seconds and a
memory limit of $3$ GB separately for solving (second column in
Table~\ref{certification_results}) and the checking of proofs and the
extraction and checking of the certificates (third column in
Table~\ref{certification_results}).

\depqbf solved and extracted proofs for 91
formulas. For about two thirds of these formulas, the tools \qbfcert and
\resqu successfully extracted and checked the certificates. These tools
implement the same approach to certificate extraction based on Q-resolution
proofs~\cite{DBLP:journals/fmsd/BalabanovJ12,DBLP:conf/sat/NiemetzPLSB12} and
show similar performance. However, the proofs produced by \depqbf had to be
converted into a different format supported by \resqu. Both \qbfcert and
\resqu represent Skolem and Herbrand functions as \emph{and-inverter graphs (AIGs)}. In contrast to
\qbfcert, the workflow of \resqu includes simplification of AIGs using the
tool \emph{ABC}~\cite{DBLP:conf/cav/BraytonM10}. We attribute the difference in the
number of instances certified by \qbfcert and \resqu to the use of ABC. \resqu
certified four instances which were not certified by \qbfcert, and \qbfcert certified
eight instances not certified by \resqu. 

In the lower part of Table~\ref{certification_results}, the workflow
implemented in \qubecert and \resqu is also based on Q-resolution proofs and
certificates in terms of Skolem and Herbrand functions and thus is most
closely related to \depqbf combined with \qbfcert and \resqu. The tool
\checker does not extract certificates, but only checks the Q-resolution proofs
produced by the solver \qubecert. The solvers \skizzo and \squolem directly
extract a certificate out of a given QBF, which is then checked using the tools
\ozziks, \qbv, or \resqu, respectively. The checker tool \ozziks is designed
to check the certificates of satisfiable instances only. Errors were reported
on 19 instances solved by \squolem when converting the extracted certificates
into the input format of \resqu. 

For the workflows that involve the extraction of Q-resolution proofs, we
observed that not only run time but also memory is critical. For example, when
solving an instance, \depqbf writes every Q-resolution step to a trace file
stored on the hard disk. This trace file is analyzed by tools like \qbfcert
and \resqu to extract a certificate. On some instances, the trace file might
become very large (up to several gigabytes) as it contains redundant
information irrelevant for the proof. The subsequent certificate
extraction step might fail due to memory limits. From the 91 instances solved
by \depqbf, proofs were extracted from the trace files for 82 instances. The
average (median) number of resolution steps in these proofs was 197,472
(2,439), ranging from one to 4,661,201 steps. For the files where the proofs
were written to, the average (median) size was 94 MB (1 MB), with a range from
0.003 MB to 1,711 MB.

Our experiments in the showcase on certificates showed that the power of
available certification workflows lags behind the power of solvers. That is
due to the fact that not every solved instance could be successfully certified
within the given time and memory limits. In practice, the size of trace files
written by solvers may hinder certification. As a remedy, solvers could
maintain resolution proofs directly in memory rather than write trace files to
the hard disk. The QRAT proof
system~\cite{DBLP:conf/cade/HeuleSB14}, for example, may be used to generate proofs in a
more compact way than Q-resolution.

%%%%%%%%%%%%%%%%%%%%%%%%%%%%%%%%%%%%%%%%%%%%%%%%%%%%%%%%%%%%%%%%%%%%%%%%%%%%%%%%
%%%%%%%%%%%%%%%%%%%%%%%%%%%%%%%%%%%%%%%%%%%%%%%%%%%%%%%%%%%%%%%%%%%%%%%%%%%%%%%%

\section{Conclusions}
\label{sec:conclusion}

We presented the experiments we conducted in the context of the 
QBF Gallery 2013, an event for the evaluation of tools related to 
QBF solving. In contrast to similar events, the QBF Gallery 2013 was 
not a competition but it was intended to be a platform for 
interested researchers to assess the state of the art of QBF technology. In the following, 
we shortly summarize our findings. 

\emph{Feedback from the Participants.} While all participants agreed that 
it is important to have a shared forum to be able to compare tools 
in a uniform setting, it turned out that the involvement of most 
participants was mainly to provide tools and fixes. There was 
a moderate discussion ongoing about benchmark selection and related
organizational matters. However, the main decisions remained with the 
organizers. Finally, most participants asked for a competition. 
The fun factor is a factor that may not be neglected, and especially 
when prizes are awarded, the motivation is even increased, 
although the event becomes more of a show than a scientific 
evaluation. 

\emph{Benchmark Selection.} We performed many experiments to 
assess the quality of the benchmark sets. We came to the conclusion 
that the selected sets sufficiently represent the benchmark collection of the QBF research community, which consists of ten thousands of formulas. 
We received new families of formulas stemming from various kinds of applications.
Overall, to compare solvers by their performance it is important that the
formulas neither are too easy such that 
all solvers can 
solve them nor too hard.  
 We made the benchmarks used available to the 
community. 

\emph{Preprocessing and Solving.} We used all tools as black boxes 
as submitted by the authors, i.e., we did not consider other options than 
the default options of the tools. Overall, we experienced that preprocessing has 
great impact on the solving performance and that the different 
preprocessors show diverse performance. Further, it turned out 
that incremental preprocessing, i.e., multiple applications of 
a preprocessor until the formula does not change anymore, 
affects solving performance in a positive way. Further, incremental 
preprocessing is more powerful than individual preprocessing
and the solvers are sensitive to the order in which 
preprocessors are applied. 
Although preprocessors usually do not implement complete decision 
procedures, often they can solve formulas directly.  
Overall, different solvers perform differently well on different 
kinds of benchmarks, i.e., the solvers are very sensitive to the 
structure of the considered problem. This indicates that
implementing a hybrid solver based on different solving paradigms 
might be promising.  
In the current experiments, we used the preprocessors in the configurations 
suggested by their authors. However,  detailed parameter tuning 
might further speed up the overall solving process. 

We found that on certain novel benchmarks ``old'' solvers 
like \nenofexm perform very well (cf.~\cite{DBLP:conf/aiia/MarinNPTG15}). Here a detailed evaluation would be 
of interest where systems available on the web are collected and 
run on recently generated encodings. Solvers like \nenofexm, \quantor, 
or \skizzo implement techniques orthogonal to approaches found in 
currently developed solvers, and explore the search space in a different manner. 
For combining old and new techniques, 
hybrid solving or portfolio approaches might be 
one promising direction of future work (cf. \cite{DBLP:journals/jair/LindauerHHS15,DBLP:journals/jsat/PulinaT10}). 

\subsection{Further Relations to Recent Advances in QBF Proof Complexity} 
The drastic differences in the performance of solvers on certain benchmark
families seem to be related to recent advances in QBF proof complexity. In the
QBF Gallery 2013 we did not run experiments to deliberately confirm
theoretical results in proof complexity. However, the global picture of our
observations to some extent appears to reflect proof theoretical properties of
approaches implemented in solvers.

Search-based solvers like \sqube and \depqbf, for example, are based on
\emph{Q-resolution}. Traditional Q-resolution~\cite{DBLP:journals/iandc/BuningKF95}
allows to resolve on existential variables and rules out tautological
clauses. \emph{Long-distance
Q-resolution}~\cite{DBLP:journals/fmsd/BalabanovJ12,DBLP:conf/iccad/ZhangM02}
generalizes Q-resolution by permitting the generation of certain tautological
resolvents. \emph{QU-resolution}~\cite{DBLP:conf/cp/Gelder12} generalizes
Q-resolution by resolving also on universal variables. The proof system
\emph{LQU$+$resolution}~\cite{DBLP:conf/sat/BalabanovWJ14} combines long-distance
Q-resolution and QU-resolution. QU-resolution and long-distance Q-resolution
were shown to be stronger than
Q-resolution~\cite{DBLP:conf/stacs/BeyersdorffCJ15,DBLP:conf/lpar/EglyLW13,DBLP:conf/cp/Gelder12}. That
is, there are classes of QBFs where any Q-resolution proof is exponentially
larger than a proof in QU- or long-distance Q-resolution. Further,
LQU$+$resolution was shown to be stronger than QU- and long-distance
Q-resolution~\cite{DBLP:conf/sat/BalabanovWJ14}.

From a practical perspective, only Q-resolution and long-distance Q-resolution
are applied for QBF solving in a systematic way. Hence the power of stronger
proof systems like LQU$+$resolution is still left unused in practice. A variant of
\depqbf and the solver
Quaffle~\cite{DBLP:conf/iccad/ZhangM02}\footnote{\url{http://www.princeton.edu/~chaff/quaffle.html}}
support long-distance Q-resolution, which however did not participate in the
QBF Gallery 2013. QU-resolution~\cite{DBLP:conf/cp/Gelder12} is
implicitly part of abstraction-based failed-literal detection for
QBF~\cite{DBLP:conf/sat/LonsingB11}, as implemented in the preprocessor
\hiqqere~\cite{DBLP:conf/sat/GelderWL12}. Hence preprocessing makes
QU-resolution available in current solving workflows. This in turn may explain
the benefits of preprocessing on solver performance, as QU-resolution is
stronger than Q-resolution, which is typically applied in search-based
solvers.

Expansion of universal variables is another successful approach to QBF solving,
in addition to backtracking search and Q-resolution.  A variant of universal
expansion was formalized as the proof system $\forall$Exp$+$Res
in~\cite{DBLP:conf/sat/JanotaM13,DBLP:journals/tcs/JanotaM15}. Thereby, initially
all universal variables are expanded. The resulting propositional formula
contains only existential variables and can be solved by Q-resolution.  The
proof system $\forall$Exp$+$Res was generalized to instantiation of universal
variables by truth constants in the proof system
IRM-calc~\cite{DBLP:conf/mfcs/BeyersdorffCJ14}. It was shown that IRM-calc
polynomially simulates the expansion-based proof system
$\forall$Exp$+$Res. That is, for any proof in $\forall$Exp$+$Res there is a
proof in IRM-calc that is at most polynomially larger.

It was shown that Q-resolution and $\forall$Exp$+$Res are incomparable with
respect to worst-case proof
sizes~\cite{DBLP:conf/stacs/BeyersdorffCJ15,DBLP:journals/tcs/JanotaM15}. That
is, there are classes of QBFs that have proofs in $\forall$Exp$+$Res of only
exponential size but Q-resolution proofs of polynomial size, and
\emph{vice versa}.
This theoretical result conforms to our observations made in the
experiments conducted in the QBF Gallery 2013. On certain instances,
expansion-based solvers clearly outperform search-based solver relying on
Q-resolution. For practical QBF solving, it may be worth combining both
expansion and Q-resolution in a single solver to benefit from the
strengths of both proof systems. So far, the generalized proof systems
IRM-calc and LQU$+$resolution~\cite{DBLP:conf/stacs/BeyersdorffCJ15} have not
been implemented in solvers and hence applying them may further improve the
state of the art.

\subsection{Outlook} 

Based on the experience gained from the QBF Gallery 2013, one year later, the follow-up event QBF Gallery 2014
 was organized in the 
context of the FLoC Olympic Games. As the 2014 edition of the Gallery 
was competitive, the organizing team was changed, because here no 
developer submitting a participating solver  should be involved in the 
organization. Many of the formulas collected for the QBF Gallery 2013
were used in the Gallery $2014$ and from the showcases, three tracks 
were derived, namely (i) the QBFLIB track, (ii) the Preprocessing track, 
and (iii) the Application track. Interestingly, all 
participants who gain benefits from using \bloqqer also submitted their 
tool with \bloqqer, what is in accordance with the license of version v35. 
In 2014, no certification track was organized, because although 
there have been several application papers in 2014 using function extraction 
facilities for solving their application problems, there are still very 
few solvers supporting certification.

\paragraph{Acknowledgments} 
We would like to thank all participants of the QBF Gallery 2013 
as well as the participants of the QBF Workshop 2013 for their 
feedback. Furthermore, we thank the anonymous reviewers for their detailed 
reports. 

%%%%%%%%%%%%%%%%%%%%%%%%%%%%%%%%%%%%%%%%%%%%%%%%%%%%%%%%%%%%%%%%%%%%%%%%%%%%%%%%
%%%%%%%%%%%%%%%%%%%%%%%%%%%%%%%%%%%%%%%%%%%%%%%%%%%%%%%%%%%%%%%%%%%%%%%%%%%%%%%%

%%%%%%%%%%%%%%%%%%%%%%%%%%%%%%%%%%%%%%%%%%%%%%%%%%%%%%%%%%%%%%%%%%%%%%%%%%%%%%%%
%%%%%%%%%%%%%%%%%%%%%%%%%%%%%%%%%%%%%%%%%%%%%%%%%%%%%%%%%%%%%%%%%%%%%%%%%%%%%%%%

\ifshowappendix

\newpage

\begin{appendix}

\section{Tables related to the Showcase on Applications (Section~\ref{sec:apps})}

\begin{table}[ht] 
\centering
 \begin{tabular}{|l|r|r|r|r|r|r|}
\hline
&\multicolumn{4}{c|}{\textbf{number of solved formulas}} & \multicolumn{2}{c|}{\textbf{runtime (sec)}}\\\hline
\textbf{solver}&solved&sat&unsat&unique&avg&total\\
\hline
\rareqs&102&37&65&5&29&46K\\
\ghostc&100&50&50&0&31&48K\\
\bghostc&100&50&50&0&32&48K\\
\nenofexm&99&56&43&0&21&48K\\
\sdooq&98&56&42&0&14&48K\\
\hiqqer&97&56&41&0&24&50K\\
\ghost&92&51&41&0&38&55K\\
\depqbfm&85&43&42&0&51&62K\\
\ldepqbf&85&43&42&0&41&62K\\
\ooqm&79&37&42&0&14&65K\\
\dooqm&79&37&42&0&14&65K\\
\sqube&60&24&36&0&66&82K\\
\hline
 \end{tabular}
 \caption{Solving statistics for the set \texttt{bomb} in \texttt{conformant-planning}.}
\label{confplanningbomb}
\end{table}

\begin{table}[ht] 
\centering
 \begin{tabular}{|l|r|r|r|r|r|r|}
\hline
&\multicolumn{4}{c|}{\textbf{number of solved formulas}} & \multicolumn{2}{c|}{\textbf{runtime (sec)}}\\\hline
\textbf{solver}&solved&sat&unsat&unique&avg&total\\
\hline
\nenofexm&149&18&131&36&64&10K\\
\hiqqer&107&18&89&0&41&43K\\
\rareqs&97&18&79&0&26&50K\\
\bghostc&92&17&75&0&73&58K\\
\ghostc&90&17&73&0&157&68K\\
\ldepqbf&85&18&67&0&15&59K\\
\depqbfm&84&17&67&0&38&62K\\
\ghost&84&16&68&0&181&74K\\
\sdooq&76&9&67&0&288&88K\\
\sqube&75&6&69&0&143&78K\\
\ooqm&74&8&66&0&356&94K\\
\dooqm&74&8&66&0&356&94K\\
\hline
 \end{tabular}
 \caption{Solving statistics for the set
   \texttt{dungeon} in \texttt{conformant-planning}.}
\label{confplanningdungeon}
\end{table}

\begin{table}[ht] 
\centering
 \begin{tabular}{|l|r|r|r|r|r|r|}
\hline
&\multicolumn{4}{c|}{\textbf{number of solved formulas}} & \multicolumn{2}{c|}{\textbf{runtime (sec)}}\\\hline
\textbf{solver}&solved&sat&unsat&unique&avg&total\\
\hline
\rareqs&149&40&109&1&30&5K\\
\nenofexm&134&39&95&0&27&18K\\
\ldepqbf&67&31&36&0&106&81K\\
\ooqm&63&29&34&0&82&83K\\
\depqbfm&60&30&30&0&102&87K\\
\dooqm&59&26&33&0&129&89K\\
\sdooq&43&27&16&0&134&102K\\
\ghost&33&19&14&0&198&111K\\
\hiqqer&32&17&15&0&82&108K\\
\bghostc&16&9&7&0&339&126K\\
\ghostc&16&9&7&0&359&126K\\
\sqube&15&10&5&0&117&123K\\
\hline
 \end{tabular}
 \caption{Solving statistics for the set \texttt{planning-CTE}.}
\label{planningCTE}
\end{table}

\begin{table}[ht] 
\centering
 \begin{tabular}{|l|r|r|r|r|r|r|}
\hline
&\multicolumn{4}{c|}{\textbf{number of solved formulas}} & \multicolumn{2}{c|}{\textbf{runtime (sec)}}\\\hline
\textbf{solver}&solved&sat&unsat&unique&avg&total\\
\hline
\ldepqbf&126&8&118&3&9&22K\\
\depqbfm&125&8&117&2&12&24K\\
\hiqqer&123&11&112&2&10&25K\\
\ghost&92&10&82&0&24&54K\\
\sdooq&91&8&83&0&47&57K\\
\ghostc&87&10&77&0&53&61K\\
\bghostc&86&10&76&0&43&61K\\
\sqube&79&8&71&0&11&64K\\
\dooqm&76&8&68&0&72&72K\\
\rareqs&49&8&41&0&52&93K\\
\ooqm&40&8&32&0&14&99K\\
\nenofexm&21&8&13&0&18&116K\\
\hline
 \end{tabular}
 \caption{Solving statistics for the set \texttt{qbf-hardness}.}
\label{qbfhardness}
\end{table}

\begin{table}[ht] 
\centering
 \begin{tabular}{|l|r|r|r|r|r|r|}
\hline
&\multicolumn{4}{c|}{\textbf{number of solved formulas}} & \multicolumn{2}{c|}{\textbf{runtime (sec)}}\\\hline
\textbf{solver}&solved&sat&unsat&unique&avg&total\\
\hline
\rareqs&121&53&68&10&23&28K\\
\hiqqer&100&50&50&0&25&47K\\
\sdooq&95&44&51&0&30&52K\\
\depqbfm&94&45&49&0&25&52K\\
\ldepqbf&94&46&48&0&31&53K\\
\ooqm&94&42&52&1&36&53K\\
\dooqm&92&41&51&0&24&54K\\
\bghostc&88&35&53&0&37&59K\\
\ghostc&83&31&52&0&39&63K\\
\sqube&77&42&35&0&16&67K\\
\ghost&67&32&35&0&31&76K\\
\nenofexm&58&29&29&0&6&83K\\
\hline
 \end{tabular}
 \caption{Solving statistics for the set \texttt{reduction-finding}.}
\label{redfinding}
\end{table}

\begin{table}[ht] 
\center
 \begin{tabular}{|l|r|r|r|r|r|r|}
\hline
&\multicolumn{4}{c|}{\textbf{number of solved formulas}} & \multicolumn{2}{c|}{\textbf{runtime (sec)}}\\\hline
\textbf{solver}&solved&sat&unsat&unique&avg&total\\
\hline
\ghostc&130&102&28&0&39&23K\\
\bghostc&130&102&28&0&40&23K\\
\ghost&116&89&27&0&40&35K\\
\hiqqer&109&79&30&3&26&39K\\
\sdooq&77&48&29&0&20&67K\\
\depqbfm&76&51&25&0&24&68K\\
\ldepqbf&75&51&24&0&19&68K\\
\sqube&59&33&26&0&38&84K\\
\rareqs&53&27&26&0&22&88K\\
\ooqm&35&6&29&0&12&103K\\
\dooqm&33&5&28&0&24&106K\\
\nenofexm&18&7&11&0&3&118K\\
\hline
 \end{tabular}
 \caption{Solving statistics for the set \texttt{sauer-reimer}.}
\label{sauerreimer}
\end{table}

\clearpage

%%%%%%%%%%%%%%%%%%%%%%%%%%%%%%%%%%%%%%%%%%%%%%%%%%%%%%%%%%%%%%%%%%%%%%%%%%%%%%%%
%%%%%%%%%%%%%%%%%%%%%%%%%%%%%%%%%%%%%%%%%%%%%%%%%%%%%%%%%%%%%%%%%%%%%%%%%%%%%%%%

\section{Tables related to the Showcase on Solving (Section~\ref{sec:solving})}
\label{sec:appendix:solving}

\begin{table}[ht]
\centering
 \begin{tabular}{|l|r|r|r|r|r|r|}

\hline
&\multicolumn{4}{c|}{\textbf{number of solved formulas}} & \multicolumn{2}{c|}{\textbf{runtime (sec)}}\\\hline
\textbf{solver}&solved&sat&unsat&unique&avg&total\\

\hline
\hiqqer&459&221&238&5&62&126K\\
\sdooq&417&190&227&0&49&156K\\
\bghostc&409&195&214&0&52&164K\\
\dooqm&409&185&224&0&43&161K\\
\ghostc&400&191&209&0&57&174K\\
\ldepqbf&395&171&224&0&37&170K\\
\depqbfm&386&167&219&0&31&176K\\
\sqube&371&161&210&0&68&202K\\
\ghost&345&172&173&0&47&217K\\
\ooqm&264&99&165&0&34&282K\\
\rareqs&258&96&162&3&25&285K\\
\nenofexm&220&106&114&7&25&318K\\
\hline
 \end{tabular}
 \caption{Solving statistics for the set \texttt{eval2010}.}
\label{eval10}
\end{table}

\begin{table}[ht]
\centering
 \begin{tabular}{|l|r|r|r|r|r|r|r|r|r|r|r|r|}
\hline

&
\begin{sideways}\sqube\end{sideways}&
\begin{sideways}\bghostc\end{sideways}&
\begin{sideways}\dooqm\end{sideways}&
\begin{sideways}\rareqs\end{sideways}&
\begin{sideways}\ghostc\end{sideways}&
\begin{sideways}\hiqqer\end{sideways}&
\begin{sideways}\ghost\end{sideways}&
\begin{sideways}\depqbfm\end{sideways}&
\begin{sideways}\nenofexm\end{sideways}&
\begin{sideways}\sdooq\end{sideways}&
\begin{sideways}\ooqm\end{sideways}&
\begin{sideways}\ldepqbf\end{sideways}\\
\hline

Ansotegui (22)
&13&5&7&7&5&13&7&11&0&12&10&12\\

Ayari (19)
&4&5&6&4&1&9&1&2&14&6&3&2\\

Biere (42)
&34&39&23&12&39&22&40&14&8&27&14&14\\

Castellini (37)
&37&37&37&37&37&37&37&37&37&37&37&37\\

Gent-Rowley (11)
&8&8&8&7&8&8&8&8&5&8&8&8\\

Herbstritt (61)
&54&37&45&41&37&45&39&53&11&48&46&54\\

Katz (3)
&3&0&0&0&0&3&0&0&1&3&0&0\\

Kontchakov (136)
&89&75&133&19&75&134&22&136&0&120&13&136\\

Letombe (52)
&50&50&50&50&50&50&51&49&41&51&50&50\\

Ling (3)
&1&3&3&3&3&3&2&3&3&3&3&3\\

Mangassarian-Veneris (23)
&6&14&13&17&13&16&8&12&17&12&13&13\\

Messinger (3)
&1&2&1&2&2&2&1&2&2&1&2&2\\

Mneimneh-Sakallah (19)
&14&18&16&0&18&10&18&0&2&13&0&0\\

Palacios (15)
&3&10&5&14&10&5&4&5&9&5&5&5\\

Pan (80)
&33&74&37&13&71&77&74&27&41&44&36&30\\

Rintanen (18)
&9&16&13&16&15&13&16&16&18&13&13&17\\

Scholl-Becker (24)
&12&16&12&16&16&12&17&11&11&14&11&12\\
\hline \hline
total (568)
&
371&
409&
409&
258&
400&
459&
345&
386&
220&
417&
264&
395\\
\hline
 \end{tabular}
 \caption{Detailed solving statistics for the set \texttt{eval2010}.}
\label{eval10suite}
\end{table}

\begin{table}[ht]
\centering
 \begin{tabular}{|l|r|r|r|r|r|r|}
\hline
solver&solved&sat&unsat&unique&avg&total\\
\hline
\ldepqbf&327&164&163&0&49&99K\\
\depqbfm&324&166&158&0&54&104K\\
\hiqqer&287&131&156&2&93&146K\\
\rareqs&250&117&133&12&26&159K\\
\sqube&227&94&133&1&117&200K\\
\dooqm&201&96&105&0&38&204K\\
\ooqm&196&97&99&0&26&206K\\
\sdooq&194&97&97&0&73&217K\\
\ghost&179&90&89&0&94&233K\\
\ghostc&179&103&76&0&79&231K\\
\bghostc&179&103&76&0&79&231K\\
\nenofexm&123&73&50&7&36&271K\\
\hline
 \end{tabular}
 \caption{Solving statistics for the set \texttt{eval2010} preprocessed with
   \bloqqer.}
\label{eval10bloqqerpp}
\end{table}

\begin{table}[ht]
\centering
 \begin{tabular}{|l|r|r|r|r|r|r|r|r|r|r|r|r|}
\hline

&
\begin{sideways}\sqube\end{sideways}&
\begin{sideways}\bghostc\end{sideways}&
\begin{sideways}\dooqm\end{sideways}&
\begin{sideways}\rareqs\end{sideways}&
\begin{sideways}\ghostc\end{sideways}&
\begin{sideways}\hiqqer\end{sideways}&
\begin{sideways}\ghost\end{sideways}&
\begin{sideways}\depqbfm\end{sideways}&
\begin{sideways}\nenofexm\end{sideways}&
\begin{sideways}\sdooq\end{sideways}&
\begin{sideways}\ooqm\end{sideways}&
\begin{sideways}\ldepqbf\end{sideways}\\
\hline

Ansotegui (22)
&13&12&7&16&12&14&14&17&3&7&11&17\\

Ayari (13)
&1&1&1&2&1&2&1&1&6&1&1&2\\

Biere (29)
&10&8&8&13&8&13&9&11&5&9&11&12\\

Castellini (27)
&27&27&27&27&27&27&27&27&27&27&27&27\\

Gent-Rowley (4)
&2&1&1&0&1&1&1&1&0&1&1&1\\

Herbstritt (50)
&37&10&38&50&10&39&11&35&1&36&38&37\\

Katz (3)
&3&1&2&1&1&3&2&3&1&2&2&3\\

Kontchakov (136)
&57&23&25&26&23&101&32&136&1&19&14&136\\

Letombe (29)
&27&26&27&28&26&25&27&26&9&27&26&26\\

Ling (3)
&3&3&3&3&3&3&3&3&3&3&3&3\\

Mangassarian-Veneris (17)
&2&9&10&12&9&10&7&10&14&9&10&9\\

Messinger (3)
&1&2&2&3&2&2&1&2&3&2&2&2\\

Mneimneh-Sakallah (17)
&13&11&16&15&11&8&6&12&5&16&14&12\\

Palacios (14)
&2&7&4&11&7&4&5&4&7&4&4&4\\

Pan (15)
&12&9&9&11&9&12&10&11&9&9&9&11\\

Rintanen (14)
&6&13&8&14&13&10&9&11&14&9&10&10\\

Scholl-Becker (24)
&11&16&13&18&16&13&14&14&15&13&13&15\\
\hline \hline
total (420)
&
227&
179&
201&
250&
179&
287&
179&
324&
123&
194&
196&
327\\
\hline
 \end{tabular}
 \caption{Detailed solving statistics for the set \texttt{eval2010}
   preprocessed with \bloqqer.}
\label{eval10_bloqqer_ppsuite}
\end{table}

\begin{table}[ht]
\begin{center}
\begin{tabular}{lrr}
\hline
       & \multicolumn{2}{c}{\emph{Number Solved}}\\
Category/& Best& Worst\\
Solver& Foot& Foot\\
\hline
\emph{NO \bloqqer (solvers perform better without \bloqqer)}&&\\
 \hiqqer & 311& 287\\
 \dooqm & 303& 141 \\
 \sdooq & 302& 194\\
 \sqube & 268& 227\\
 \bghostc & 262& 180\\
 \ghostc & 262& 180\\
 \ghost & 204& 179\\
 \ooqm & 163& 142\\
\hline
\emph{WANT \bloqqer (solvers perform better with \bloqqer)} &&\\
 \ldepqbf & 327& 303\\
 \depqbfm & 323 & 297\\
 \rareqs & 250& 184\\
 \nenofexm & 121& 115\\
\hline
\end{tabular}
\end{center}
\caption{Classification of solvers into two categories depending on their
performance on 568 instances of the set \texttt{eval2010} with (category
``WANT \bloqqer'') and without prior preprocessing by \bloqqer (category ``NO
\bloqqer''). In each of these categories, column ``Best Foot'' shows the
numbers of instances that were solved when choosing to run on preprocessed
instances or on original ones, respectively. Column ``Worst Foot'', on the
contrary, shows the numbers of instances solved when making the opposite choice.
}
\label{bestfooteval2010RECOMPUTED}
\end{table}

%%%%%%%%%%%%%%%%%%%%%%%%%%%%%%%%%%%%%%%%%%%%%%%%%%%%%%%%%%%%%%%%%%%%%%%%%%%%%%%%
%%%%%%%%%%%%%%%%%%%%%%%%%%%%%%%%%%%%%%%%%%%%%%%%%%%%%%%%%%%%%%%%%%%%%%%%%%%%%%%%

\end{appendix}
\fi

%%%%%%%%%%%%%%%%%%%%%%%%%%%%%%%%%%%%%%%%%%%%%%%%%%%%%%%%%%%%%%%%%%%%%%%%%%%%%%%%
%%%%%%%%%%%%%%%%%%%%%%%%%%%%%%%%%%%%%%%%%%%%%%%%%%%%%%%%%%%%%%%%%%%%%%%%%%%%%%%%

\end{document}

%% file: macros.tex
\newcommand{\sqube}{\textsf{QuBE}\xspace}
\newcommand{\bghostc}{\textsf{bGhostQ-CEGAR}\xspace}
\newcommand{\dooqm}{\textsf{dual$\_$Ooq}\xspace}
\newcommand{\rareqs}{\textsf{RAReQS}\xspace}

\newcommand{\ghostc}{\textsf{GhostQ-CEGAR}\xspace}
\newcommand{\hiqqer}{\textsf{Hiqqer3}\xspace}
\newcommand{\ghost}{\textsf{GhostQ}\xspace}
\newcommand{\depqbfm}{\textsf{DepQBF}\xspace}
\newcommand{\nenofexm}{\textsf{Nenofex}\xspace}
\newcommand{\quantor}{\textsf{Quantor}\xspace}
\newcommand{\sdooq}{\textsf{sDual$\_$Ooq}\xspace}
\newcommand{\ooqm}{\textsf{Qoq}\xspace}
\newcommand{\ldepqbf}{\textsf{DepQBF-lazy-qpup}\xspace}
\newcommand{\resqu}{\textsf{ResQu}\xspace}
\newcommand{\qbfcert}{\textsf{QBFcert}\xspace}
\newcommand{\freetoqbf}{\textsf{free2qbf}\xspace}
\newcommand{\minitoqbf}{\textsf{mini2qbf}\xspace}

\newcommand{\skizzo}{\textsf{sKizzo}\xspace}
\newcommand{\ozziks}{\textsf{ozziKs}\xspace}
\newcommand{\squolem}{\textsf{squolem}\xspace}
\newcommand{\qbv}{\textsf{qbv}\xspace}
\newcommand{\qubecert}{\textsf{QuBE-cert}\xspace}
\newcommand{\checker}{\textsf{checker}\xspace}

\newcommand{\squeezebf}{\textsf{sQueezeBF}\xspace}
\newcommand{\depqbf}{\textsf{DepQBF}\xspace}
\newcommand{\bloqqer}{\textsf{Bloqqer}\xspace}
\newcommand{\hiqqerp}{\textsf{Hiqqer3p}\xspace}
\newcommand{\hiqqere}{\textsf{Hiqqer3e}\xspace}

\newcommand{\preproA}{\textsf{Hiqqer3e}\xspace}
\newcommand{\preproB}{\textsf{Bloqqer}\xspace}
\newcommand{\preproC}{\textsf{Hiqqer3p}\xspace}
\newcommand{\preproD}{\textsf{SqueezeBF}\xspace}